\documentclass[prl,superscriptaddress,twocolumn,10pt,nofootinbib,longbibliography,floatfix]{revtex4-2}
\usepackage{amssymb}
\usepackage{amsmath}
\usepackage{appendix}
\usepackage{times}
\usepackage{graphicx}
\usepackage{subeqnarray}
\usepackage{epsfig}
\usepackage{epstopdf}

\usepackage{color}
\usepackage{graphics}\usepackage{epsfig} 
\usepackage{multirow}
\usepackage{bbding}
\usepackage[normalem]{ulem}
\useunder{\uline}{\ul}{}
\begin{document}

\title{Magnonic Quantum Spin Hall Effect with Chiral Magnon Transport in Bilayer Altermagnets}

\author{Bo Yuan}
\author{Yingxi Bai}
\author{Ying Dai}
\email{daiy60@sdu.edu.cn}
\author{Baibiao Huang}
\author{Chengwang Niu}
\email{c.niu@sdu.edu.cn}
\affiliation
{School of Physics, State Key Laboratory of Crystal Materials, Shandong University, 250100 Jinan, China}

\begin{abstract}

Altermagnetism has attracted considerable interest, yet its associated spintronic phenomena have so far been largely confined to electronic systems. In this work, we uncover a universal symmetry-based strategy for realizing topological altermagnets with the magnonic quantum spin Hall effect, as evidenced by a nonzero spin Chern number and protected helical edge states. Moreover, we demonstrate that chiral magnon splitting in altermagnets gives rise to an intrinsically anisotropic, momentum-resolved thermal Hall response, sharply contrasting with those in ferromagnets and antiferromagnets, thus offering enhanced flexibility for selective manipulation. As a concrete material realization, first-principles calculations and Heisenberg-DM model analysis reveal that V$_2$WS$_4$ bilayer exhibits $d$-wave altermagnetism, integer spin Chern number with helical magnon edge states, and the nonzero momentum-locked thermal Hall conductivity. Our results establish a direct link between topological magnons and altermagnetism, opening new avenues for dissipationless magnonic devices.

\end{abstract}
\maketitle
\date{\today}

\vspace{0.2cm}

\noindent{M}agnons, as the quanta of spin waves, are fundamental quasiparticles in magnetically ordered materials~\cite{Nat.Rev.Mater.61114}. Benefiting from their charge neutrality and the absence of Joule heating, magnons support almost dissipationless wave-like information transfer, launching the field of magnonicss~\cite{AFMe17690,Nat.Phys.201511453}. Specifically, the observation of magnon thermal Hall effect in Lu$_{2}$V$_{2}$O$_{7}$ has motivated the extension of topological band theory to magnonic systems~\cite{observation}. A conceptual milestone is the topological magnon insulator (TMI), which has been experimentally confirmed in ferromagnets (FM) CrI$_3$, CrSiTe$_3$, and Mn$_5$Ge$_3$ with the nontrivial topology characterized by a nonzero Chern number and single-chiral propagating edge state~\cite{SNE1,SNE2,SNE3,SNE4,SNE5,THE1,THE2,THE3,THE4,DMI1,DMI2}. Remarkably, for TMI in antiferromagnets (AFM), the antiparallel spins give rise to the pseudo–time-reversal symmetry, which enforces degenerate magnon branches with opposite chiralities, constituting the bosonic analog of the quantum spin Hall effect, characterized by a nonzero spin Chern number and a pair of helical edge states~\cite{PTRS1,PTRS2,QSH1,QSH2}. However, realizing a TMI with the magnonic quantum spin Hall effect remains elusive to date. Moreover, the vanishing net magnetization in antiferromagnets leads to multiple advantages, such as ultrafast magnon dynamics, exceptional tolerance to external magnetic fields, and the absence of stray fields, thereby offering strong prospects for low-dissipation spin transport and ultrafast information processing~\cite{127077202,202505779,Optically,Antif}.

Only recently, altermagnet (AM) has attracted significant interest, propelling progress in both fundamental physics and spintronic technologies~\cite{al1,al3,al4,al5,thermal,spinpolar}. As a newly established category of collinear magnetic order, it describes the collinear antiparallel ordering without net magnetization but alternating reciprocal-space spin polarization, synergistically combining the advantages of FM and AFM~\cite{al2,alter3,alter4,alter5,alter7,PhysRevX.12.011028}. Research on altermagnetism is advancing rapidly, fueled both by the continual identification of new candidate materials and by the key experimental verification in established systems ~\cite{NP2025cry,PhysRevLett.133.206401,ex1,ex2}. Indeed, altermagnetism provides new opportunities to explore phenomena such as the anomalous Hall effect, spin-polarized transport, and magneto-optical responses~\cite{133146602,AM36.29,PhysRevLett.134.106802,PhysRevLett.134.106801,zbbr-426l}. Moreover, AM supports a distinct class of THz-range magnons, which exhibit strong coupling between exchange modes of opposite chirality—a feature absent in conventional AFM~\cite{alter7,gn6c-1q19}. And the anisotropic and chirality-selective magnon decay can be obtained in altermagnets, offering opportunities for symmetry-driven chiral magnon propagation and nonreciprocal magnon transport~\cite{b5vs-ldpm,chiral}. However, despite significant progress in the experimental observation of chiral magnonic band splitting, topological magnons in altermagnets remain scarce~\cite{7yhz-jptc}. Crucially, a general strategy to construct TMI with nonzero spin Chern numbers and helical edge states is in high demand to bridge altermagnetism and band topology, which is essential for advancing the field of magnonics.

In this work, focusing on square bilayer AM, which possess chiral magnon band splitting, we propose a rational design principle of the TMI with magnonic quantum spin Hall effect. By incorporating spin-group symmetries, we demonstrate that even when pseudo–time-reversal symmetry is broken, the helical edge states in AM remain protected and are characterized by a nontrivial spin Chern number of $C_{s} = \pm1$, due to the combined effects of interlayer pseudospin flipping and the intrinsic rotational symmetry of the altermagnetic order. Furthermore, based on an analysis of site-symmetry groups, we predict specific Wyckoff positions within square layer groups that can simultaneously support altermagnetism and helical edge states. Remarkably, it is found that AM hosts exotic magnon transport characteristics distinct from those in both FM and AFM configurations, resulting in the nonzero momentum-resolved thermal Hall conductivity (MTHC) that exhibits momentum- and chirality-locked directional behavior. For material realization, combined with first-principles calculations and a Heisenberg-DM model, we uncover that the V$_2$WS$_4$ bilayer exhibits AM chiral magnon splitting and hosts topologically nontrivial magnon states with nonzero $C_{s}$, helical edge states, and momentum-resolved magnon transport. This work not only pinpoints candidate materials supporting magnonic quantum spin Hall effect in AM, but also establishes universal design principles, offering a versatile platform for the discovery and engineering of topological magnons.

\begin{figure}
	\centering
	\includegraphics[width=250pt]{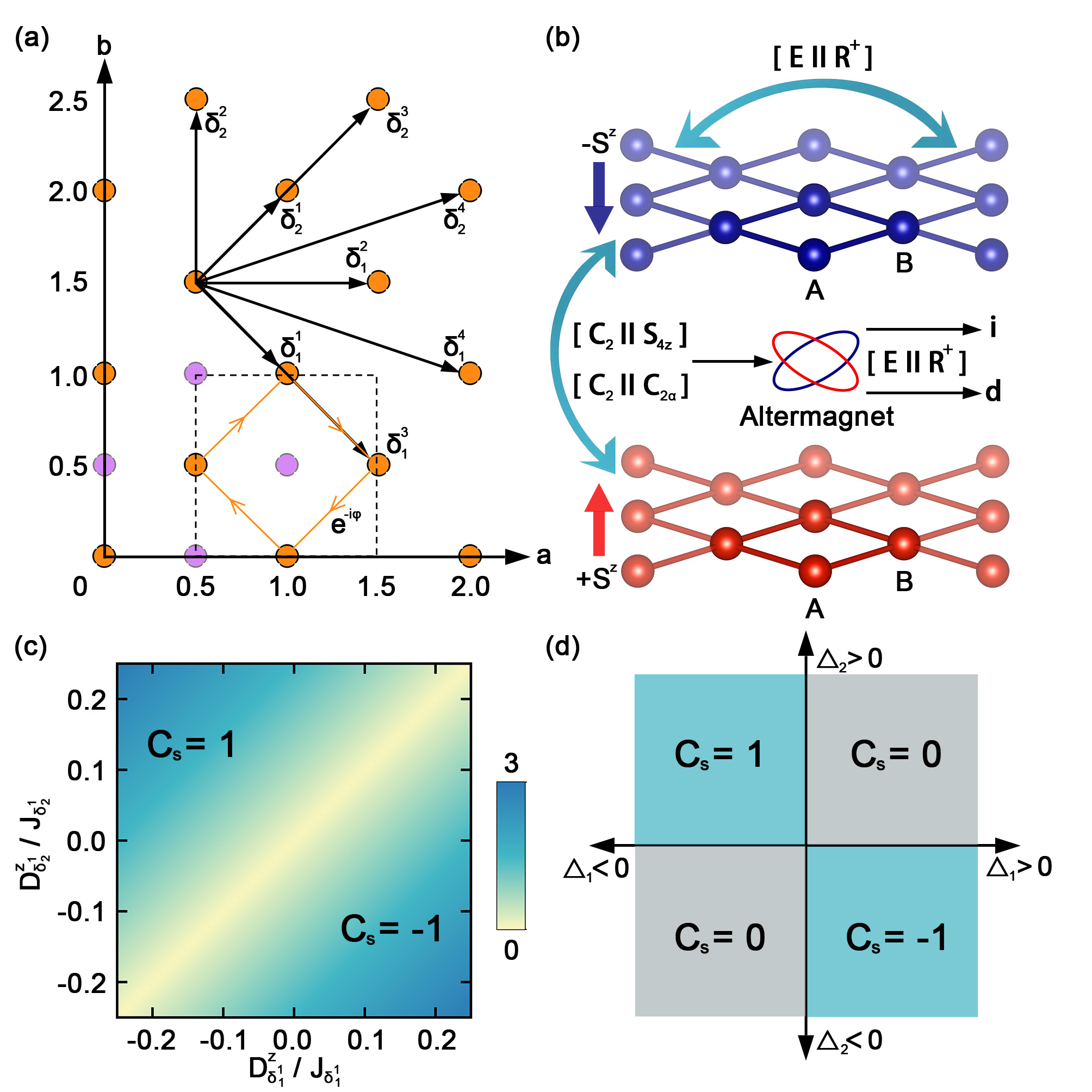}
	\caption{(a) Distribution of two checkerboard sublattices (orange and purple circles), highlighting an inequivalent closed loop $\mathcal{O}$ (solid orange lines) within a unit cell (black dashed lines), together with the directions of magnetic interactions up to $\mathcal{N}=4$. (b) Schematic illustration of the origin of altermagnetism in the square bilayer and the associated intra- and interlayer symmetry operations. The topological phase diagrams versus (c) $D_{\delta_{1}^{1}}^{z}/J_{\delta_{1}^{1}}$ and $D_{\delta_{2}^{1}}^{z}/J_{\delta_{2}^{1}}$, and (d) $\Delta_{1}$ and $\Delta_{2}$, defined as $\Delta_{1}=J_{\delta_{1}^{2}}^{A}-J_{\delta_{1}^{2}}^{B}$ and $\Delta_{2}=J_{\delta_{2}^{2}}^{A}-J_{\delta_{2}^{2}}^{B}$, respectively.} 
	\label{Fig.3.}
\end{figure} 	

\noindent{{\large\textbf{Results}}}

\noindent{\textbf{Magnonic quantum spin Hall effect in altermagnet}}

\noindent{T}o elucidate the mechanisms underlying the alternating chiral band splitting and the emergence of nontrivial TMI with magnonic quantum spin Hall effect in square lattices, we construct a fundamental spin Hamiltonian with layered out-of-plane collinear magnetic order, which can be expressed as
\begin{equation}\label{Eq4}
	\begin{aligned}
		H\!=&\!\sum_{l,(i,j)\in \mathcal{N}}\!J_{\delta_{1/2}^{\mathcal{N}}}^{l}(\boldsymbol{S_{i}^{l}\!\cdot S_{j}^{l}})\!+\!\sum_{l,\mathcal{N} = 1}\!D_{\delta_{1/2}^{\mathcal{N}}}^{l}\boldsymbol{z}\cdot(\boldsymbol{S_{i}^{l}\!\times S_{j}^{l}})\\
		&+\!\sum_{i,l,l^{'}}\!J_{\perp}(\boldsymbol{S_{i}^{l}\!\cdot \!S_{i}^{l^{'}}})\!+\!\sum_{i,l}\!A_{z}(S_{i,z}^{l})^{2}.
	\end{aligned}
\end{equation}
Where $\boldsymbol{S_{i}^{l}}$ indicates the spin operator at lattice site $i$ in layer $l =~1,2$. $J_{\delta_{1/2}^{\mathcal{N}}}^{l}$ denotes the Heisenberg exchange interaction, with $\delta_{1/2}^{\mathcal{N}}$ labeling the two $\{S_{4z},C_{2\alpha}\}$-related directions for the $\mathcal{N}$th-neighbor couplings, as illustrated in Fig.~\ref{Fig.3.}(a). $S_{4z}$ is a fourfold rotation combined with a mirror symmetry with respect to the $z$ axis, and $C_{2\alpha}$ is a twofold rotation along the direction $[\cos(\alpha), \sin(\alpha), 0]$. $D_{\delta_{1/2}^{\mathcal{N}}}^{l}$ represents the Dzyaloshinskii-Moriya interaction (DMI). Moreover, in the third and fourth terms, we define $J_{\perp}$ and $A_{z}$ as the interlayer exchange interaction and easy-axis anisotropy, respectively. By performing the Holstein–Primakoff (HP) transformation  and fourier transformation (see Supplementary Note 1), the above spin Hamiltonian can be recast into $H=\sum_{k}\Psi_{\boldsymbol{k}}^{\dagger}H_{k}\Psi_{\boldsymbol{k}}$ with
\begin{equation}\label{k}
	H_{k}=\left[
	\begin{array}{cccc}
		h_{1}(\boldsymbol{k}) & 0 & 0 & h_{\perp}\\
		0 & h_{2}(\boldsymbol{k}) & h_{\perp} & 0\\
		0 & h_{\perp}^{\dagger} & h_{1}^{T}(-\boldsymbol{k}) & 0\\
		h_{\perp}^{\dagger} & 0 & 0 & h_{2}^{T}(-\boldsymbol{k})
	\end{array}
	\right],	
\end{equation}
and the basis is $\Psi_{\boldsymbol{k}} = (\psi_{\boldsymbol{k}},\psi_{\boldsymbol{-k}}^{\dagger})^{T}$. Here,  $\psi_{\boldsymbol{k}} = (b_{A,\boldsymbol{k}}^{1},b_{B,\boldsymbol{k}}^{1},b_{A,\boldsymbol{k}}^{2},b_{B,\boldsymbol{k}}^{2})^{T}$ with 1 (2) labels the layer index and A (B) represents the sublattice index as annotated on Fig.~\ref{Fig.3.}(b). $h_{\perp}$ denotes the interlayer interaction and $h_{l}(\boldsymbol{k})=h_{l,0}\sigma_{0}+\sum_{i = x,y,z}h_{l,i}\sigma_{i}$. $\sigma_{0}$ indicates the unit matrix and $\sigma_{i}$ are the Pauli matrices. 

To investigate the nonrelativistic magnonic band features, we first exclude the DMI with $h_{l,y} = 0$. In the framework of spin groups, the symmetry operations can be formally written as $[R_i \Vert R_j]$. $R_i$ acts solely on the spin space and $R_j$ acts solely on the real space. The operations $[R_{i}\vert\vert R_{j}]$ can be classified into two categories for the A-type magnetic order depending on whether $R_{j}$ reverses the $z$-direction, denoted as $[E\vert\vert R^{+}]$ and $[C_{2}\vert\vert R^{-}]$, with $R^{+}$ and $R^{-}$ contains $\{E, C_{nz}, m_{\alpha}\}$ and $\{\bar{E}, m_{z}, C_{2\alpha}, S_{nz}\}$, respectively. Therefore, we obtain $[E\Vert R^{+}] = \tau_{0}s_{0}\sigma_{0/x}$ and $[C_{2}\Vert R^{-}] = \tau_{x}s_{x}\sigma_{0/x}$ due to coupling between the layer degree of freedom and the spin degree of freedom, and $\sigma_{0/x}$ denotes whether the operation exchanges the A and B sublattices. Under the symmetry operation $P=[C_{2}\vert\vert R^{-}]$, which satisfies $PH_{k}P^{-1}=H_{P^{-1}k}$ for $P=[C_{2}\vert\vert R^{-}]$, the Hamiltonian components must obey $h_{1}(\boldsymbol{k}) = h^{*}_{2}(P\boldsymbol{k})$. In particular, for $P=[C_{2}\vert\vert \bar{E}]$, imposing $h_{1}(\boldsymbol{k}) = h^{*}_{2}(-\boldsymbol{k})$ endows the system with pseudo–time-reversal symmetry, i.e., $\Theta \sum_{z}H_{k} = \sum_{z}H_{-k}\Theta$ with $\Theta = i\sigma_{z}s_{y}\sigma_{0}K$ and $K$ denotes the complex conjugation, resulting in “Kramers degeneracy" for the magnons. Furthermore, the same result can be induced by $[C_{2}\vert\vert M_{z}]$ symmetry, as it acts in conjunction with the  $[E\vert\vert C_{2z}]$ symmetry of the square lattice. Consequently, we identify seven layer groups with $R^{-}\in \{S_{4z}, C_{2\alpha }\}$ ($\alpha=0$ or $\frac{\pi}{4}$) possible for altermagnetism, as summarized in Table~\ref{table} I.

\begin{table}[]\label{table}
	\caption{Seven layer groups capable of hosting altermagnetism in a bilayer square lattice with $S_{4z}$ and/or $C_{2\alpha}$ symmetry, together with the Wyckoff positions of the checkerboard lattice and their corresponding site-symmetry groups. Magnon chiral splitting is induced in various layer groups only upon inclusion of magnetic exchange interactions up to at least the $\mathcal{N}$th-nearest neighbor. The specific Wyckoff positions that stabilize helical magnon edge modes with spin Chern number $|C_{s}|=1$ are predicted across these layer groups.} 
	\begin{tabular}{|c|c|c|c|c|}
		\hline
		{\ul }            Layer group&Wyckoff Position  & \ $\mathcal{N}$ \ &Site-symmetry group  & $|C_{s}|=1$ \\ \hline
		\multirow{2}{*}{}50 &$2c + 2d$ & 1 & $2..$ &  \CheckmarkBold  \\ \cline{2-5} 
		$P\bar{4}$& $2e + 2e'$ & 1 & $2..$ &  \CheckmarkBold   \\ \hline
		\multirow{2}{*}{}53 & $2d+2e$ & 4 & $4..$ & \XSolidBold \\ \cline{2-5} 
		$P422$& $4f$ & 4 & 2.. &   \CheckmarkBold \\ \hline
		\multirow{2}{*}{} 54& $2b+2b'$ & 4 & $4..$ & \XSolidBold \\ \cline{2-5} 
		$P42_{1}2$& $4c$ & 1 & $2..$ &  \CheckmarkBold \\ \hline
		\multirow{2}{*}{}57 & $2d+2e$ & 1 & $2.mm$ & \XSolidBold \\ \cline{2-5} 
		$P\bar{4}2m$& $4h$ & 1 & $2..$ &  \CheckmarkBold \\ \hline
		\multirow{2}{*}{} 58& $2b+2b'$ & 1 & $2.mm$  & \XSolidBold  \\ \cline{2-5} 
		$P\bar{4}2_{1}m$& $4c$ &  1& $2..$ &  \CheckmarkBold \\ \hline
		\multirow{2}{*}{}59 & $2c+2d$ & 2 & $2mm.$ &  \CheckmarkBold \\ \cline{2-5} 
		$P\bar{4}m2$& $2e+2e'$ & 2 & $2mm.$ &  \CheckmarkBold \\ \hline
		\multirow{2}{*}{} 60& $4c$ & 1 & $2..$ &   \CheckmarkBold \\ \cline{2-5} 
		$P\bar{4}b2$& $4d$ & 1 & $2..$ &  \CheckmarkBold \\ \hline
	\end{tabular}
	
\end{table}

Moreover, the altermagnetic waveform is restricted by the symmetry operation $[E\vert\vert R^{+}]$. For two distinct directions $\delta_{1}^{\mathcal{N}}$ and $\delta_{2}^{\mathcal{N}}$ that are related by $[C_{2}\vert\vert S_{4z}, C_{2\alpha}]$ symmetry as plotted in Fig.~\ref{Fig.3.}(a), the corresponding exchange interaction yields no contribution to the magnon chiral splitting in altermagnet if they are safeguarded by $[E\vert\vert R^{+}]$ symmetry. For instance, under the $[E\vert\vert C_{4z}]$ symmetry, exchange interactions along the perpendicular axes $\delta_{1}^{\mathcal{N}}$ and $\delta_{2}^{\mathcal{N}}$ do not induce chiral magnon splitting for $\mathcal{N}=1$–3. Only at $\mathcal{N}=4$ does such splitting become allowed, giving rise to $i$-wave altermagnetism, as listed in Table~\ref{table} I. In all other configurations, $d$-wave altermagnetism is observed. When the $[E\vert\vert M_{\alpha=0}]$ symmetry is preserved, magnon chiral splitting can only be realized at least by taking into account the next-nearest-neighbor ($\mathcal{N}=2$) magnetic exchange interactions.

To explore the topological properties, the antisymmetric out-of-plane DMI $\boldsymbol{D}_{\delta_{1/2}^{1}}$ $\vert\vert$ $\boldsymbol{z}$ is considered, which generates a finite phase factor $e^{-i\varphi_{\langle ij \rangle} }$ with $\varphi_{\langle ij \rangle} = atan(D_{\delta_{1/2}^{1}}^{z}/J_{\delta_{1/2}^{1}})$. Remarkably, a fictitious magnetic flux $\phi = \sum_{\mathcal{O}}\varphi_{\langle ij \rangle}$ emerges by summing these phases along the closed $\mathcal{O}$ in the two distinct checkerboard configurations correspond to magnetic atoms positioned at (0, 0, $\pm$z), (0.5, 0.5, $\pm$z) or (0.5, 0, $\pm$z), (0, 0.5, $\pm$z) as shown in Fig.\ref{Fig.3.}(a)~\cite{PhysRevLett.61.2015, thermal}. As illustrated in Figs.~\ref{Fig.3.}(c) and~\ref{Fig.3.}(d), when $D_{\delta_{1}^{1}}^{z}/J_{\delta_{1}^{1}}$ $\neq$ $D_{\delta_{2}^{1}}^{z}/J_{\delta_{2}^{1}}$ and $\Delta_{1}$ and $\Delta_{2}$ share opposite signs, this flux leads to the nonzero spin Chern noumber $C_{s} = \pm1$. Here, $\Delta_{1}=J_{\delta_{1}^{2}}^{A}-J_{\delta_{1}^{2}}^{B}$, $\Delta_{2}=J_{\delta_{2}^{2}}^{A}-J_{\delta_{2}^{2}}^{B}$, and the spin Chern noumber is defined as $C_{s} =(C_{L}+C_{R})/2$, with $C_{L(R)}=\frac{1}{2 \pi}\int_{Bz}\Omega_{L(R)}(\boldsymbol{k})d^{2}\boldsymbol{k}$ are the Chern numbers for left-handed and right-handed magnonic bands, respectively. Therefore, the emergence of TMI phase requires the breaking of both $C_{4z}$ and $M_{\alpha=\frac{\pi}{4}}$ symmetries at the site of magnetic atoms, since $C_{4z}$ enforces $\phi = 0$ with $D_{\delta_{1}^{1}}^{z}/J_{\delta_{1}^{1}} = D_{\delta_{2}^{1}}^{z}/J_{\delta_{2}^{1}}$, whereas $M_{\alpha=\frac{\pi}{4}}$ symmetry constrains $\Delta_{1}$ and $\Delta_{2}$ to have the same sign.

\begin{figure}
	\centering
	\includegraphics[width=250pt]{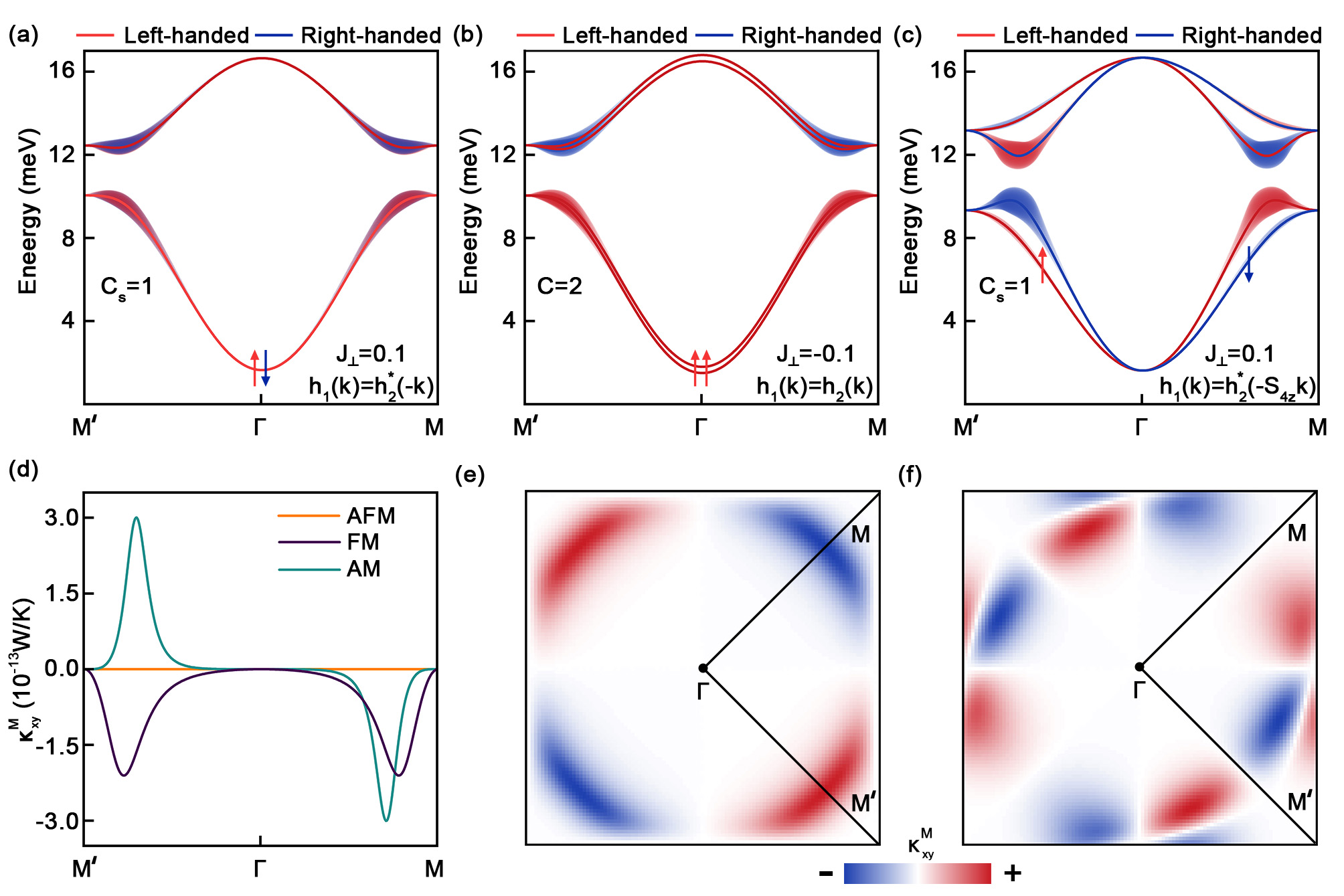}
	\caption{Magnon band structures of Heisenberg-DM model under (a) AFM, (b) FM, and (c) AM configurations, with the fat band analysis weighted with the magnonic Berry curvatures. (d) The distribution of momentum-resolved thermal Hall conductivity $\kappa^{M}_{xy}(50, \boldsymbol{k})$ along M$^{\prime}$-$\Gamma$-M for AFM, FM, and AM configurations at 50$K$. The distribution of $\kappa^{M}_{xy}(50, \boldsymbol{k})$ in whole Brillouin zone for AM with (e) $d$-wave  and (f) $i$-wave configurations.} 
	\label{Fig.4.}
\end{figure} 	

\noindent{\textbf{Chiral magnon transport in altermagnet}}

\noindent{W}e then show how altermagnetism gives rise to distinctive topological magnon transport properties, fundamentally different from those in FM and AFM systems. The corresponding magnon band structures for the three magnetic orders are separately plotted Figs.~\ref{Fig.4.}(a-c). In the AFM configuration, the combined $\bar{E}T$ symmetry enforces $h_{1}(\boldsymbol{k}) = h^{*}_{2}(-\boldsymbol{k})$, preserving the degeneracy of magnon bands with opposite chirality. Consequently, the Berry curvatures satisfy $\Omega_{L}(\boldsymbol{k})=-\Omega_{R}(\boldsymbol{k})$, leading to a nontrivial topological phase with $C_{s}=1$. In the FM configuration, where $h_{1}(\boldsymbol{k}) = h_{2}(\boldsymbol{k})$, the magnon bands realize a TMI phase with Chern number $C = 2$, as the magnonic bands below the gap carry Berry curvatures of the same sign. In the case of altermagnetism, taking $d$-wave configuration at $4f$ sites in layer group $P\bar{4}2_{1}m$ as an example, the breaking of $\bar{E}T$ symmetry leads to chiral splitting of the magnon bands. Meanwhile, the $S_{4z}T$ symmetry protects symmetric splittings with opposite chirality along the $M^{\prime}$–$\Gamma$ and $M$–$\Gamma$ directions, enforcing the relation $\Omega_{L}(\boldsymbol{k}) = -\Omega_{R}(S_{4z}\boldsymbol{k})$. Consequently, the anisotropic chiral magnons in AM offer an extra degree of freedom for controlling magnon transport, which motivates the introduction of the concept of momentum-resolved thermal Hall conductivity (MTHC)
\begin{equation}
	\begin{aligned}
		\kappa^{M}_{xy}(T, \boldsymbol{k})=-\frac{k_{B}^{2}T}{(2\pi)^{2}\hbar}\sum_{n}c_{2}(n_{B}(\epsilon))\Omega_{n}^{xy}(\boldsymbol{k}),\\
	\end{aligned}
\end{equation}
where $k_{B}$ and $T$ are the Boltzmann constant and temperature, respectively. $n_{B}(\epsilon)$ represents the Bose distribution function, defined as $n_{B}(\epsilon)=(e^{\frac{\epsilon-\mu}{k_{B}T}}-1)^{-1}$. $c_{2}(\tau)=(1+\tau)ln^{2}[(1+\tau)/\tau]-ln^{2}\tau-2Li_{2}(-\tau)$ with $Li_{2}$ denotes the polylogarithm function. In Fig.~\ref{Fig.4.}(d), we present the path distribution of $\kappa^{M}_{xy}$ at $T = 50 K$ for the three magnetic configurations, evaluated along the direction of altermagnetic magnon band splitting. Owing to the $\bar{E}T$–enforced degeneracy of chiral magnon bands, $\kappa^{M}_{xy}$ vanishes in the AFM configuration, while finite values emerge in both FM and AM phases. Notably, compared with the FM case, MTHC in AM displays the strongly momentum-dependent and directionally anisotropic response. For a more detailed characterization of $\kappa^{M}_{xy}$, projection maps over the full Brillouin zone for the AM and FM configurations are plotted (see Figs.~\ref{Fig.4.}(e) and Supplementary Fig.1), respectively. The response of $\kappa^{M}_{xy}$ to the wave vector direction is completely isotropic in FM, thus yielding a non-zero net macroscopic response corresponding to $\kappa_{xy}(T)=\int_{BZ}\kappa^{M}_{xy}(T, \boldsymbol{k})d\boldsymbol{k}$. In the AM phase, the presence of $S_{4z}T$ symmetry enforces a $d$-wave–symmetric distribution of $\kappa^{M}_{xy}$, directly reflecting the momentum-locking characteristic of chiral magnon splitting. Moreover, the $i$-wave AM supports a momentum-locked $\kappa^{M}_{xy}$ whose spatial distribution follows an $i$-wave symmetry, as shown in Fig.~\ref{Fig.4.}(f).

\begin{figure}
	\centering
	\includegraphics[width=250pt]{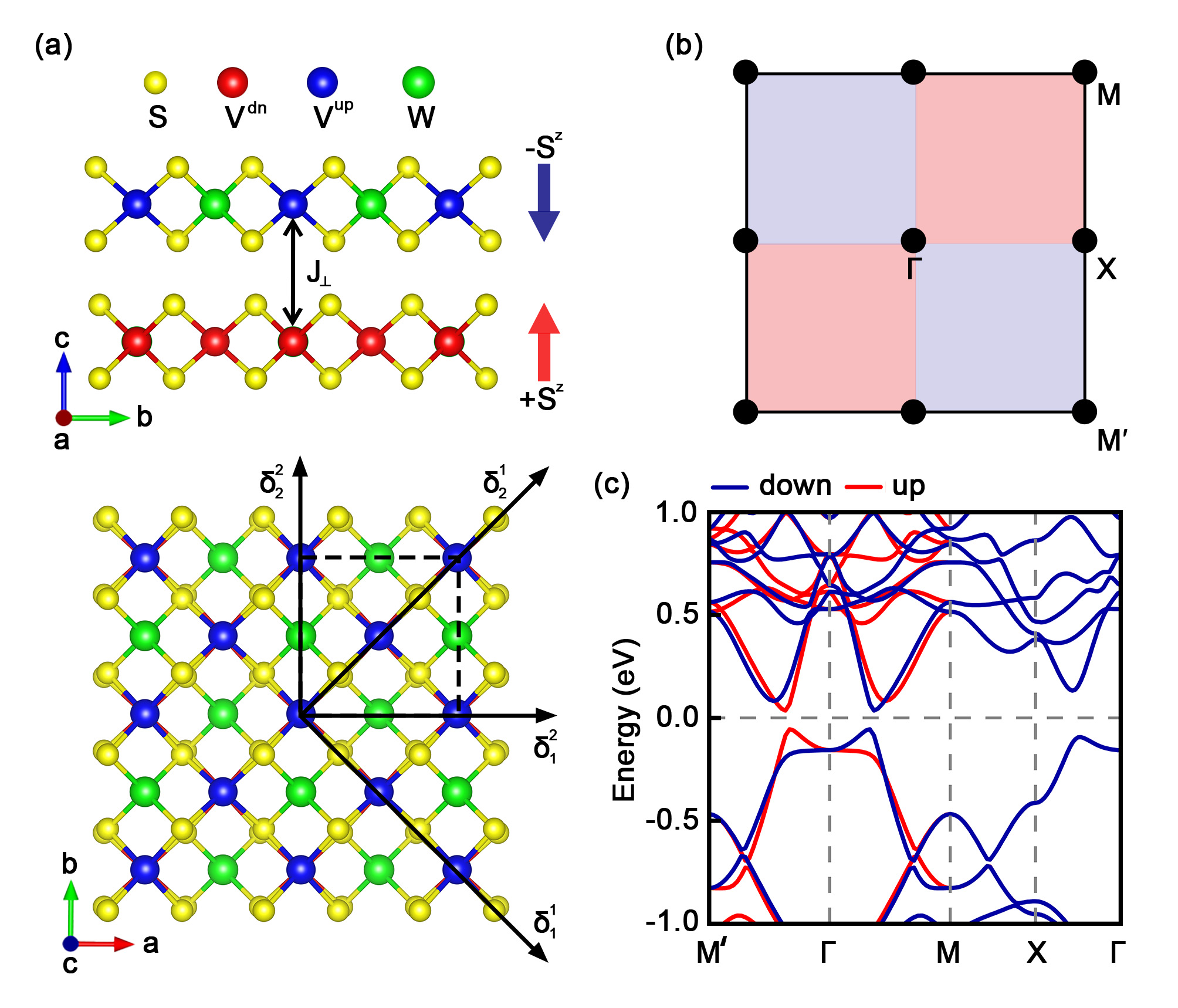}
	\caption{(a) Top and side views of the $V_{2}WS_{4}$ bilayer in layer group $P\bar{4}2_{1}m$. (b) The first Brillouin zone with marked high symmetry points. (c) Electronic band structure of the $V_{2}WS_{4}$ bilayer without spin-orbit coupling. The alternating reciprocal-space spin polarization is clearly visible.}
	\label{Fig.1.}
\end{figure}

\noindent{\textbf{Material realization}}

\noindent{T}hen, based on density functional theory and the Heisenberg-DM model, we demonstrate that the V$_2$WS$_4$ bilayer exhibits altermagnetic chiral magnon splitting and hosts nontrivial topological magnon states with helical edge modes. The V$_{2}$WS$_{4}$ bilayer has a square lattice, belonging to the layer group $P\bar{4}2_{1}m$ (No.58). The fully optimized lattice constant is a = 5.74~\AA. Total energy calculations indicate that the bilayer favors a spin-polarized state, with the magnetic moment primarily localized on the V atoms and having a magnitude of 2.6 $\mu_{B}$. Each primitive cell contains four V atoms, located at the $4c$ Wyckoff positions, with site symmetry group $2..$, as illustrated in Fig.~\ref{Fig.1.}(a). The V$_{2}$WS$_{4}$ bilayer stabilizes an out-of-plane magnetic ground state, featuring parallel intralayer ordering and antiparallel interlayer coupling, which intriguingly leads to zero net magnetization. Figure~\ref{Fig.1.}(c) presents the band structure of the V$_{2}$WS$_{4}$ bilayer without SOC. Clearly, the bands show spin degeneracy along the M–X–$\Gamma$ path but undergo spin splitting along $\Gamma$–M and $\Gamma$–M$'$, where the two directions host opposite spin polarizations. Such momentum-dependent band spin splitting is a hallmark of altermagnetism.

\begin{figure}
	\centering
	\includegraphics[width=250pt]{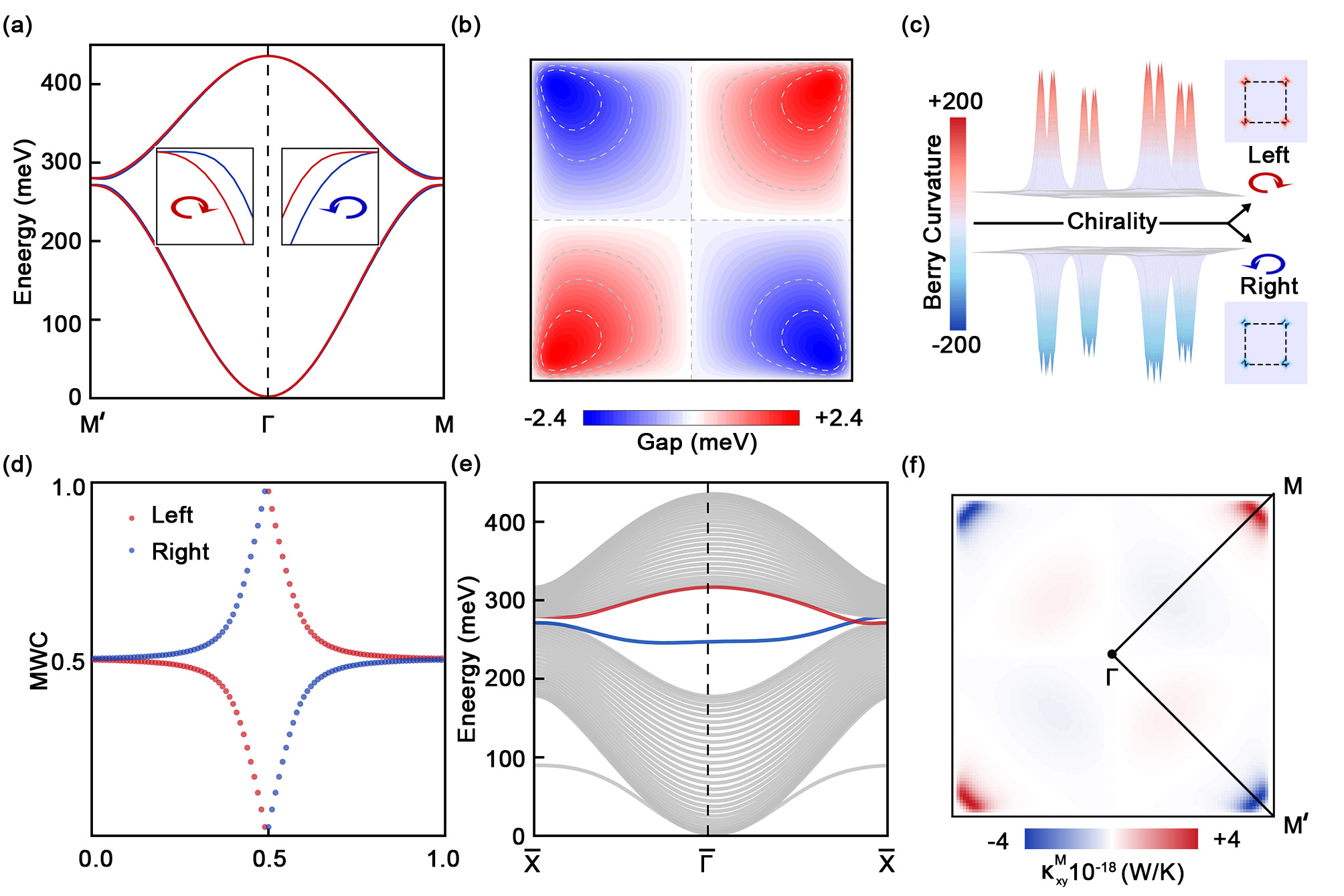}
	\caption{(a) Magnonic band structure of the $V_{2}WS_{4}$ bilayer. Red and blue colors mark the opposite magnon chiralities. (b) Spin-splitting energy projections given by the energy difference $E_{(L, k)} - E_{_{(R, k)}}$. (c) Distribution of magnonic Berry curvatures of the two bands with opposite chiralities below the gap for $V_{2}WS_{4}$ bilayer. (d) Evolution of magnonic Wannier centers (MWC) for the $V_{2}WS_{4}$ bilayer, indicating a nonzero spin Chern number of $C_{s} = 1$. (e) Magnonic helical edge states of the $V_{2}WS_{4}$ bilayer. (f) The distribution of $\kappa^{M}_{xy}(200, \boldsymbol{k})$ in whole Brillouin zone for $V_{2}WS_{4}$ bilayer.}
	\label{Fig.2.}
\end{figure}

To elucidate the magnon excitation properties of the V$_2$WS$_4$ bilayer, the magnetic exchange interactions are extracted using the four-state methodology (4SM) and are presented in Table S1. Figure~\ref{Fig.2.}(a) presents the magnonic band structure of the V$_{2}$WS$_{4}$~\cite{4sm}. The interlayer pseudospin flipping generates two magnonic branches with opposite chirality, namely the left-handed mode $E_{(L,k)}$ and right-handed mode $E_{(R,k)}$, which correspond to $S_{z}=\pm 1$, respectively. As illustrated in Fig.~\ref{Fig.2.}(b), the chiral band splitting acquires a $d$-wave–type distribution, arising from the interplay between the $S_{4z}$ symmetry and interlayer spin-flip mechanisms. In fact, the two chiral modes satisfy the relation $E_{(R,k)} = E_{(L,S_{4z}k)}$. Notably, the antisymmetric DMI opens a magnonic band gap in both chiral branches. The topologically nontrivial nature of the gap can be explicitly confirmed through the calculation of the spin Chern number $C_{s}$. Figure~\ref{Fig.2.}(c) shows the distribution of Berry curvatures $\Omega_{L(R)}(\boldsymbol{k})$ for magnonic bands with opposite chirality. We see that, the dominant contribution to the Berry curvature originates from the vicinity of M, and its opposite signs for the two chiralities lead to Chern numbers $C_L$ = 1 and $C_R$ = -1 for the respective modes. This is further demonstrated by our calculations of the magnonic Wannier centers as plotted in Fig.~\ref{Fig.2.}(d). Thus, the spin Chern number is $C_{s}$ = 1, indicating that the V$_{2}$WS$_{4}$ bilayer with altermagnetic ordering is a TMI. A defining characteristic of a TMI with $C_{s}$ = 1 is the emergence of one pair of helical edge states inside the insulating magnonic band gap. To illustrate this hallmark, we evaluate the magnonic edge states of a nanoribbon, with the resulting dispersion shown in Fig.~\ref{Fig.2.}(e). As expected, the emergence of one pair of helical edge states in the nanoribbons is clearly visible, in direct agreement with the nonzero spin Chern number $C_{s}$ = 1. In addition, the emergence of nonzero magnonic Berry curvature together with $d$-wave altermagnetism implies the existence of the momentum-locked $\kappa^{M}_{xy}$. As illustrated in Fig.~\ref{Fig.2.}(f), $\kappa^{M}_{xy}$ exhibits a pronounced $d$-wave–symmetric distribution in momentum space, directly reflecting the chiral magnon splitting and highlighting the intimate connection between altermagnetic symmetry and directional thermal Hall transport. Moreover, we calculate the integral of $\kappa^{M}_{xy}$ around the momentum $M$ point and obtain a value on the order of $10^{-14}$ W/K, implying that a detectable thermal Hall conductivity can be realized in altermagnets by selectively exciting magnons at specific momenta.

\noindent{{\large\textbf{Conclusion}}}

\noindent{I}n conclusion, we have identified definitive symmetry and structural criteria required to realize square altermagnets with chiral magnon band splitting and topologically nontrivial helical edge states. On the one hand, the presence of an $R^{-} \in \{ S_{4z}, C_{2\alpha}\}$ symmetry that exchanges two oppositely spin-polarized sublattices between layers is required, which constrains the system to seven specific layer groups. On the other hand, the magnetic atoms are required to form an inequivalent closed loop within the unit cell, which fixes their Wyckoff positions, and their site-symmetry group must exclude both the $C_{4z}$ and $M_{\alpha=\pi/4}$ symmetries. Concurrently, we uncovered distinctive transport behavior in altermagnets, where a MTHC emerges with tunable chirality and transverse deflection governed by the spin-wave momentum. Furthermore, we identified the V$_{2}$WS$_{4}$ bilayer as a viable material realization of the proposed TMI characterized by nonzero spin Chern number $C_{s} = 1$ and helical edge states. Our findings not only advance magnon topology in unconventional magnetism but also offer new insights into transport engineering mediated by altermagnetism.

\noindent{{\large\textbf{Methods}}}

\noindent{\textbf{Calculation}}

\noindent{W}e perform first-principles calculations for structural relaxations and static calculations using density functional theory as implemented in the Vienna ab initio simulation package (VASP), employing the Perdew-Burke-Ernzerhof (PBE) generalized gradient approximation (GGA) for the exchange-correlation potential~\cite{vasp,PBE}. A 30~\AA~ vacuum layer is introduced between slabs for AB-stacked $V_{2}WS_{4}$ in VASP simulations to eliminate inter-slab interactions. We enforced convergence thresholds of 0.001 eV/\AA~ for ionic relaxation and $10^{-6}$ for electronic self-consistency, while maintaining a 500 eV plane-wave cutoff energy to ensure sufficient basis set completeness for both wavefunctions and pseudopotentials. We employ the GGA+U method using Hubbard parameters U = 4 eV for V and U = 1 eV for W, specifically targeting d-orbitals. Exchange interactions and DMI are determined via the 4SM in a 3×3×1 supercell configuration.

\noindent{\textbf{HP transformation in the linear spin-wave theory}}

\noindent{W}ithin the framework of linear spin-wave theory, the Holstein-Primakoff transformation maps spin operators to bosonic creation and annihilation operators, which are subsequently represented in a matrix form: ${\bf S}_{i}$ = $\hat{M}_{i}{\bf a}_{i}$ with 
\begin{equation}\label{k}
	\hat{M}_{i}=\frac{\sqrt{2S}}{2}\left[
	\begin{array}{ccc}
		1 & 1 & 0 \\
		-i & i & 0\\
		0 & 0 & \sqrt{\frac{2}{S}}
	\end{array}
	\right],
	{\bf a}_{i}=\left[
\begin{array}{c}
	a_{i}\\
	a_{i}^{\dagger}\\
	S_{i}-a_{i}^{\dagger}a_{i}
\end{array}
\right].
\end{equation}

In collinear systems, the opposite spin ${\bf S}_{i}^{'}$ can be represnted as ${\bf S}_{i}^{'}= \hat{R} {\bf S}_{i}$ with 
\begin{equation}\label{k}
	\hat{R}=\left[
	\begin{array}{ccc}
		-1 & 0 & 0 \\
		0 & 1 & 0\\
		0 & 0 & -1
	\end{array}
	\right].
\end{equation}

\acknowledgments{\noindent{T}his work was supported by the Key R$\&$D Program of Shandong Province, China (Grant No.2025CXPT204), National Natural Science Foundation of China (Grant No. 12174220), Shandong Provincial Science Foundation for Excellent Young Scholars (Grant No. ZR2023YQ001), and Taishan Scholar Program of Shandong Province.}


\begin{thebibliography}{52}%
	\makeatletter
	\providecommand \@ifxundefined [1]{%
		\@ifx{#1\undefined}
	}%
	\providecommand \@ifnum [1]{%
		\ifnum #1\expandafter \@firstoftwo
		\else \expandafter \@secondoftwo
		\fi
	}%
	\providecommand \@ifx [1]{%
		\ifx #1\expandafter \@firstoftwo
		\else \expandafter \@secondoftwo
		\fi
	}%
	\providecommand \natexlab [1]{#1}%
	\providecommand \enquote  [1]{``#1''}%
	\providecommand \bibnamefont  [1]{#1}%
	\providecommand \bibfnamefont [1]{#1}%
	\providecommand \citenamefont [1]{#1}%
	\providecommand \href@noop [0]{\@secondoftwo}%
	\providecommand \href [0]{\begingroup \@sanitize@url \@href}%
	\providecommand \@href[1]{\@@startlink{#1}\@@href}%
	\providecommand \@@href[1]{\endgroup#1\@@endlink}%
	\providecommand \@sanitize@url [0]{\catcode `\\12\catcode `\$12\catcode
		`\&12\catcode `\#12\catcode `\^12\catcode `\_12\catcode `\%12\relax}%
	\providecommand \@@startlink[1]{}%
	\providecommand \@@endlink[0]{}%
	\providecommand \url  [0]{\begingroup\@sanitize@url \@url }%
	\providecommand \@url [1]{\endgroup\@href {#1}{\urlprefix }}%
	\providecommand \urlprefix  [0]{URL }%
	\providecommand \Eprint [0]{\href }%
	\providecommand \doibase [0]{https://doi.org/}%
	\providecommand \selectlanguage [0]{\@gobble}%
	\providecommand \bibinfo  [0]{\@secondoftwo}%
	\providecommand \bibfield  [0]{\@secondoftwo}%
	\providecommand \translation [1]{[#1]}%
	\providecommand \BibitemOpen [0]{}%
	\providecommand \bibitemStop [0]{}%
	\providecommand \bibitemNoStop [0]{.\EOS\space}%
	\providecommand \EOS [0]{\spacefactor3000\relax}%
	\providecommand \BibitemShut  [1]{\csname bibitem#1\endcsname}%
	\let\auto@bib@innerbib\@empty
	%</preamble>
	\bibitem [{\citenamefont {Pirro}\ \emph {et~al.}(2021)\citenamefont {Pirro},
		\citenamefont {Vasyuchka}, \citenamefont {Serga},\ and\ \citenamefont
		{Hillebrands}}]{Nat.Rev.Mater.61114}%
	\BibitemOpen
	\bibfield  {author} {\bibinfo {author} {\bibfnamefont {P.}~\bibnamefont
			{Pirro}}, \bibinfo {author} {\bibfnamefont {V.~I.}\ \bibnamefont
			{Vasyuchka}}, \bibinfo {author} {\bibfnamefont {A.~A.}\ \bibnamefont
			{Serga}},\ and\ \bibinfo {author} {\bibfnamefont {B.}~\bibnamefont
			{Hillebrands}},\ }\bibfield  {title} {\bibinfo {title} {Advances in coherent
			magnonics},\ }\href {https://doi.org/10.1038/s41578-021-00332-w} {\bibfield
		{journal} {\bibinfo  {journal} {Nat. Rev. Mater.}\ }\textbf {\bibinfo
			{volume} {6}},\ \bibinfo {pages} {1114} (\bibinfo {year} {2021})}\BibitemShut
	{NoStop}%
	\bibitem [{\citenamefont {Wang}\ \emph {et~al.}()\citenamefont {Wang},
		\citenamefont {Xia}, \citenamefont {Xu}, \citenamefont {Lan}, \citenamefont
		{Han},\ and\ \citenamefont {Yu}}]{AFMe17690}%
	\BibitemOpen
	\bibfield  {author} {\bibinfo {author} {\bibfnamefont {Y.}~\bibnamefont
			{Wang}}, \bibinfo {author} {\bibfnamefont {J.}~\bibnamefont {Xia}}, \bibinfo
		{author} {\bibfnamefont {H.}~\bibnamefont {Xu}}, \bibinfo {author}
		{\bibfnamefont {G.}~\bibnamefont {Lan}}, \bibinfo {author} {\bibfnamefont
			{X.}~\bibnamefont {Han}},\ and\ \bibinfo {author} {\bibfnamefont
			{G.}~\bibnamefont {Yu}},\ }\bibfield  {title} {\bibinfo {title} {Magnons in
			van der waals antiferromagnetic materials},\ }\href
	{https://doi.org/https://doi.org/10.1002/adfm.202517690} {\bibfield
		{journal} {\bibinfo  {journal} {Adv. Funct. Mater.}\ }\textbf {\bibinfo
			{volume} {n/a}},\ \bibinfo {pages} {e17690}}\BibitemShut {NoStop}%
	\bibitem [{\citenamefont {Chumak}\ \emph {et~al.}()\citenamefont {Chumak},
		\citenamefont {Vasyuchka}, \citenamefont {Serga},\ and\ \citenamefont
		{Hillebrands}}]{Nat.Phys.201511453}%
	\BibitemOpen
	\bibfield  {author} {\bibinfo {author} {\bibfnamefont {A.~V.}\ \bibnamefont
			{Chumak}}, \bibinfo {author} {\bibfnamefont {V.~I.}\ \bibnamefont
			{Vasyuchka}}, \bibinfo {author} {\bibfnamefont {A.~A.}\ \bibnamefont
			{Serga}},\ and\ \bibinfo {author} {\bibfnamefont {B.}~\bibnamefont
			{Hillebrands}},\ }\bibfield  {title} {\bibinfo {title} {Magnon spintronics},\
	}\href {https://doi.org/10.1038/nphys3347} {\bibfield  {journal} {\bibinfo
			{journal} {Nat. Phys.}\ }\textbf {\bibinfo {volume} {11}},\ \bibinfo {pages}
		{453}}\BibitemShut {NoStop}%
	\bibitem [{\citenamefont {Onose}\ \emph {et~al.}(2010)\citenamefont {Onose},
		\citenamefont {Ideue}, \citenamefont {Katsura}, \citenamefont {Shiomi},
		\citenamefont {Nagaosa},\ and\ \citenamefont {Tokura}}]{observation}%
	\BibitemOpen
	\bibfield  {author} {\bibinfo {author} {\bibfnamefont {Y.}~\bibnamefont
			{Onose}}, \bibinfo {author} {\bibfnamefont {T.}~\bibnamefont {Ideue}},
		\bibinfo {author} {\bibfnamefont {H.}~\bibnamefont {Katsura}}, \bibinfo
		{author} {\bibfnamefont {Y.}~\bibnamefont {Shiomi}}, \bibinfo {author}
		{\bibfnamefont {N.}~\bibnamefont {Nagaosa}},\ and\ \bibinfo {author}
		{\bibfnamefont {Y.}~\bibnamefont {Tokura}},\ }\bibfield  {title} {\bibinfo
		{title} {Observation of the magnon {H}all effect},\ }\href
	{https://doi.org/10.1126/science.1188260} {\bibfield  {journal} {\bibinfo
			{journal} {Science}\ }\textbf {\bibinfo {volume} {329}},\ \bibinfo {pages}
		{297} (\bibinfo {year} {2010})}\BibitemShut {NoStop}%
	\bibitem [{\citenamefont {Cheng}\ \emph {et~al.}(2016)\citenamefont {Cheng},
		\citenamefont {Okamoto},\ and\ \citenamefont {Xiao}}]{SNE1}%
	\BibitemOpen
	\bibfield  {author} {\bibinfo {author} {\bibfnamefont {R.}~\bibnamefont
			{Cheng}}, \bibinfo {author} {\bibfnamefont {S.}~\bibnamefont {Okamoto}},\
		and\ \bibinfo {author} {\bibfnamefont {D.}~\bibnamefont {Xiao}},\ }\bibfield
	{title} {\bibinfo {title} {Spin {N}ernst effect of magnons in collinear
			antiferromagnets},\ }\href {https://doi.org/10.1103/PhysRevLett.117.217202}
	{\bibfield  {journal} {\bibinfo  {journal} {Phys. Rev. Lett.}\ }\textbf
		{\bibinfo {volume} {117}},\ \bibinfo {pages} {217202} (\bibinfo {year}
		{2016})}\BibitemShut {NoStop}%
	\bibitem [{\citenamefont {Go}\ and\ \citenamefont {Kim}(2022)}]{SNE2}%
	\BibitemOpen
	\bibfield  {author} {\bibinfo {author} {\bibfnamefont {G.}~\bibnamefont
			{Go}}\ and\ \bibinfo {author} {\bibfnamefont {S.~K.}\ \bibnamefont {Kim}},\
	}\bibfield  {title} {\bibinfo {title} {Tunable large spin {N}ernst effect in
			a two-dimensional magnetic bilayer},\ }\href
	{https://doi.org/10.1103/PhysRevB.106.125103} {\bibfield  {journal} {\bibinfo
			{journal} {Phys. Rev. B}\ }\textbf {\bibinfo {volume} {106}},\ \bibinfo
		{pages} {125103} (\bibinfo {year} {2022})}\BibitemShut {NoStop}%
	\bibitem [{\citenamefont {Kovalev}\ and\ \citenamefont {Zyuzin}(2016)}]{SNE3}%
	\BibitemOpen
	\bibfield  {author} {\bibinfo {author} {\bibfnamefont {A.~A.}\ \bibnamefont
			{Kovalev}}\ and\ \bibinfo {author} {\bibfnamefont {V.}~\bibnamefont
			{Zyuzin}},\ }\bibfield  {title} {\bibinfo {title} {Spin torque and {N}ernst
			effects in {D}zyaloshinskii-{M}oriya ferromagnets},\ }\href
	{https://doi.org/10.1103/PhysRevB.93.161106} {\bibfield  {journal} {\bibinfo
			{journal} {Phys. Rev. B}\ }\textbf {\bibinfo {volume} {93}},\ \bibinfo
		{pages} {161106} (\bibinfo {year} {2016})}\BibitemShut {NoStop}%
	\bibitem [{\citenamefont {Zhang}\ and\ \citenamefont {Cheng}(2022)}]{SNE4}%
	\BibitemOpen
	\bibfield  {author} {\bibinfo {author} {\bibfnamefont {H.}~\bibnamefont
			{Zhang}}\ and\ \bibinfo {author} {\bibfnamefont {R.}~\bibnamefont {Cheng}},\
	}\bibfield  {title} {\bibinfo {title} {A perspective on magnon spin {N}ernst
			effect in antiferromagnets},\ }\bibfield  {journal} {\bibinfo  {journal}
		{Appl. Phys. Lett.}\ }\textbf {\bibinfo {volume} {120}},\ \href
	{https://doi.org/10.1063/5.0084359} {10.1063/5.0084359} (\bibinfo {year}
	{2022})\BibitemShut {NoStop}%
	\bibitem [{\citenamefont {Cui}\ \emph {et~al.}(2023)\citenamefont {Cui},
		\citenamefont {Zeng}, \citenamefont {Cui}, \citenamefont {Yu},\ and\
		\citenamefont {Yang}}]{SNE5}%
	\BibitemOpen
	\bibfield  {author} {\bibinfo {author} {\bibfnamefont {Q.}~\bibnamefont
			{Cui}}, \bibinfo {author} {\bibfnamefont {B.}~\bibnamefont {Zeng}}, \bibinfo
		{author} {\bibfnamefont {P.}~\bibnamefont {Cui}}, \bibinfo {author}
		{\bibfnamefont {T.}~\bibnamefont {Yu}},\ and\ \bibinfo {author}
		{\bibfnamefont {H.}~\bibnamefont {Yang}},\ }\bibfield  {title} {\bibinfo
		{title} {Efficient spin {S}eebeck and spin {N}ernst effects of magnons in
			altermagnets},\ }\href {https://doi.org/10.1103/PhysRevB.108.L180401}
	{\bibfield  {journal} {\bibinfo  {journal} {Phys. Rev. B}\ }\textbf {\bibinfo
			{volume} {108}},\ \bibinfo {pages} {L180401} (\bibinfo {year}
		{2023})}\BibitemShut {NoStop}%
	\bibitem [{\citenamefont {Zhang}\ \emph {et~al.}(2021)\citenamefont {Zhang},
		\citenamefont {Xu}, \citenamefont {Carnahan}, \citenamefont {Sretenovic},
		\citenamefont {Suri}, \citenamefont {Xiao},\ and\ \citenamefont {Ke}}]{THE1}%
	\BibitemOpen
	\bibfield  {author} {\bibinfo {author} {\bibfnamefont {H.}~\bibnamefont
			{Zhang}}, \bibinfo {author} {\bibfnamefont {C.}~\bibnamefont {Xu}}, \bibinfo
		{author} {\bibfnamefont {C.}~\bibnamefont {Carnahan}}, \bibinfo {author}
		{\bibfnamefont {M.}~\bibnamefont {Sretenovic}}, \bibinfo {author}
		{\bibfnamefont {N.}~\bibnamefont {Suri}}, \bibinfo {author} {\bibfnamefont
			{D.}~\bibnamefont {Xiao}},\ and\ \bibinfo {author} {\bibfnamefont
			{X.}~\bibnamefont {Ke}},\ }\bibfield  {title} {\bibinfo {title} {Anomalous
			thermal {H}all effect in an insulating van der waals magnet},\ }\href
	{https://doi.org/10.1103/PhysRevLett.127.247202} {\bibfield  {journal}
		{\bibinfo  {journal} {Phys. Rev. Lett.}\ }\textbf {\bibinfo {volume} {127}},\
		\bibinfo {pages} {247202} (\bibinfo {year} {2021})}\BibitemShut {NoStop}%
	\bibitem [{\citenamefont {Hirschberger}\ \emph {et~al.}(2015)\citenamefont
		{Hirschberger}, \citenamefont {Chisnell}, \citenamefont {Lee},\ and\
		\citenamefont {Ong}}]{THE2}%
	\BibitemOpen
	\bibfield  {author} {\bibinfo {author} {\bibfnamefont {M.}~\bibnamefont
			{Hirschberger}}, \bibinfo {author} {\bibfnamefont {R.}~\bibnamefont
			{Chisnell}}, \bibinfo {author} {\bibfnamefont {Y.~S.}\ \bibnamefont {Lee}},\
		and\ \bibinfo {author} {\bibfnamefont {N.~P.}\ \bibnamefont {Ong}},\
	}\bibfield  {title} {\bibinfo {title} {Thermal {H}all effect of spin
			excitations in a kagome magnet},\ }\href
	{https://doi.org/10.1103/PhysRevLett.115.106603} {\bibfield  {journal}
		{\bibinfo  {journal} {Phys. Rev. Lett.}\ }\textbf {\bibinfo {volume} {115}},\
		\bibinfo {pages} {106603} (\bibinfo {year} {2015})}\BibitemShut {NoStop}%
	\bibitem [{\citenamefont {Katsura}\ \emph {et~al.}(2010)\citenamefont
		{Katsura}, \citenamefont {Nagaosa},\ and\ \citenamefont {Lee}}]{THE3}%
	\BibitemOpen
	\bibfield  {author} {\bibinfo {author} {\bibfnamefont {H.}~\bibnamefont
			{Katsura}}, \bibinfo {author} {\bibfnamefont {N.}~\bibnamefont {Nagaosa}},\
		and\ \bibinfo {author} {\bibfnamefont {P.~A.}\ \bibnamefont {Lee}},\
	}\bibfield  {title} {\bibinfo {title} {Theory of the thermal {H}all effect in
			quantum magnets},\ }\href {https://doi.org/10.1103/PhysRevLett.104.066403}
	{\bibfield  {journal} {\bibinfo  {journal} {Phys. Rev. Lett.}\ }\textbf
		{\bibinfo {volume} {104}},\ \bibinfo {pages} {066403} (\bibinfo {year}
		{2010})}\BibitemShut {NoStop}%
	\bibitem [{\citenamefont {Zhou}\ \emph {et~al.}(2022)\citenamefont {Zhou},
		\citenamefont {Liu}, \citenamefont {Wu}, \citenamefont {Jiang}, \citenamefont
		{Shi}, \citenamefont {Li}, \citenamefont {Sui}, \citenamefont {Hu},\ and\
		\citenamefont {Luo}}]{THE4}%
	\BibitemOpen
	\bibfield  {author} {\bibinfo {author} {\bibfnamefont {X.}~\bibnamefont
			{Zhou}}, \bibinfo {author} {\bibfnamefont {H.}~\bibnamefont {Liu}}, \bibinfo
		{author} {\bibfnamefont {W.}~\bibnamefont {Wu}}, \bibinfo {author}
		{\bibfnamefont {K.}~\bibnamefont {Jiang}}, \bibinfo {author} {\bibfnamefont
			{Y.}~\bibnamefont {Shi}}, \bibinfo {author} {\bibfnamefont {Z.}~\bibnamefont
			{Li}}, \bibinfo {author} {\bibfnamefont {Y.}~\bibnamefont {Sui}}, \bibinfo
		{author} {\bibfnamefont {J.}~\bibnamefont {Hu}},\ and\ \bibinfo {author}
		{\bibfnamefont {J.}~\bibnamefont {Luo}},\ }\bibfield  {title} {\bibinfo
		{title} {Anomalous thermal {H}all effect and anomalous {N}ernst effect of
			{C}s{V}$_{3}${S}b$_{5}$},\ }\href
	{https://doi.org/10.1103/PhysRevB.105.205104} {\bibfield  {journal} {\bibinfo
			{journal} {Phys. Rev. B}\ }\textbf {\bibinfo {volume} {105}},\ \bibinfo
		{pages} {205104} (\bibinfo {year} {2022})}\BibitemShut {NoStop}%
	\bibitem [{\citenamefont {Moriya}(1960)}]{DMI1}%
	\BibitemOpen
	\bibfield  {author} {\bibinfo {author} {\bibfnamefont {T.}~\bibnamefont
			{Moriya}},\ }\bibfield  {title} {\bibinfo {title} {New mechanism of
			anisotropic superexchange interaction},\ }\href@noop {} {\bibfield  {journal}
		{\bibinfo  {journal} {Phys. Rev. Lett.}\ }\textbf {\bibinfo {volume} {4}},\
		\bibinfo {pages} {228} (\bibinfo {year} {1960})}\BibitemShut {NoStop}%
	\bibitem [{\citenamefont {Kundu}\ and\ \citenamefont {Zhang}(2015)}]{DMI2}%
	\BibitemOpen
	\bibfield  {author} {\bibinfo {author} {\bibfnamefont {A.}~\bibnamefont
			{Kundu}}\ and\ \bibinfo {author} {\bibfnamefont {S.}~\bibnamefont {Zhang}},\
	}\bibfield  {title} {\bibinfo {title} {Dzyaloshinskii-{M}oriya interaction
			mediated by spin-polarized band with {R}ashba spin-orbit coupling},\ }\href
	{https://doi.org/10.1103/PhysRevB.92.094434} {\bibfield  {journal} {\bibinfo
			{journal} {Phys. Rev. B}\ }\textbf {\bibinfo {volume} {92}},\ \bibinfo
		{pages} {094434} (\bibinfo {year} {2015})}\BibitemShut {NoStop}%
	\bibitem [{\citenamefont {Li}\ \emph {et~al.}(2022)\citenamefont {Li},
		\citenamefont {Wu}, \citenamefont {Luo}, \citenamefont {Huang},\ and\
		\citenamefont {Chang}}]{PTRS1}%
	\BibitemOpen
	\bibfield  {author} {\bibinfo {author} {\bibfnamefont {Y.-M.}\ \bibnamefont
			{Li}}, \bibinfo {author} {\bibfnamefont {Y.-J.}\ \bibnamefont {Wu}}, \bibinfo
		{author} {\bibfnamefont {X.-W.}\ \bibnamefont {Luo}}, \bibinfo {author}
		{\bibfnamefont {Y.}~\bibnamefont {Huang}},\ and\ \bibinfo {author}
		{\bibfnamefont {K.}~\bibnamefont {Chang}},\ }\bibfield  {title} {\bibinfo
		{title} {Higher-order topological phases of magnons protected by magnetic
			crystalline symmetries},\ }\href
	{https://doi.org/10.1103/PhysRevB.106.054403} {\bibfield  {journal} {\bibinfo
			{journal} {Phys. Rev. B}\ }\textbf {\bibinfo {volume} {106}},\ \bibinfo
		{pages} {054403} (\bibinfo {year} {2022})}\BibitemShut {NoStop}%
	\bibitem [{\citenamefont {Kondo}\ \emph {et~al.}(2019)\citenamefont {Kondo},
		\citenamefont {Akagi},\ and\ \citenamefont {Katsura}}]{PTRS2}%
	\BibitemOpen
	\bibfield  {author} {\bibinfo {author} {\bibfnamefont {H.}~\bibnamefont
			{Kondo}}, \bibinfo {author} {\bibfnamefont {Y.}~\bibnamefont {Akagi}},\ and\
		\bibinfo {author} {\bibfnamefont {H.}~\bibnamefont {Katsura}},\ }\bibfield
	{title} {\bibinfo {title} {${Z}_{2}$ topological invariant for magnon spin
			{H}all systems},\ }\href {https://doi.org/10.1103/PhysRevB.99.041110}
	{\bibfield  {journal} {\bibinfo  {journal} {Phys. Rev. B}\ }\textbf {\bibinfo
			{volume} {99}},\ \bibinfo {pages} {041110} (\bibinfo {year}
		{2019})}\BibitemShut {NoStop}%
	\bibitem [{\citenamefont {Kane}\ and\ \citenamefont {Mele}(2005)}]{QSH1}%
	\BibitemOpen
	\bibfield  {author} {\bibinfo {author} {\bibfnamefont {C.~L.}\ \bibnamefont
			{Kane}}\ and\ \bibinfo {author} {\bibfnamefont {E.~J.}\ \bibnamefont
			{Mele}},\ }\bibfield  {title} {\bibinfo {title} {Quantum spin {H}all effect
			in graphene},\ }\href {https://doi.org/10.1103/PhysRevLett.95.226801}
	{\bibfield  {journal} {\bibinfo  {journal} {Phys. Rev. Lett.}\ }\textbf
		{\bibinfo {volume} {95}},\ \bibinfo {pages} {226801} (\bibinfo {year}
		{2005})}\BibitemShut {NoStop}%
	\bibitem [{\citenamefont {Qi}\ and\ \citenamefont {Zhang}(2011)}]{QSH2}%
	\BibitemOpen
	\bibfield  {author} {\bibinfo {author} {\bibfnamefont {X.-L.}\ \bibnamefont
			{Qi}}\ and\ \bibinfo {author} {\bibfnamefont {S.-C.}\ \bibnamefont {Zhang}},\
	}\bibfield  {title} {\bibinfo {title} {Topological insulators and
			superconductors},\ }\href {https://doi.org/10.1103/RevModPhys.83.1057}
	{\bibfield  {journal} {\bibinfo  {journal} {Rev. Mod. Phys.}\ }\textbf
		{\bibinfo {volume} {83}},\ \bibinfo {pages} {1057} (\bibinfo {year}
		{2011})}\BibitemShut {NoStop}%
	\bibitem [{\citenamefont {Bossini}\ \emph {et~al.}(2021)\citenamefont
		{Bossini}, \citenamefont {Pancaldi}, \citenamefont {Soumah}, \citenamefont
		{Basini}, \citenamefont {Mertens}, \citenamefont {Cinchetti}, \citenamefont
		{Satoh}, \citenamefont {Gomonay},\ and\ \citenamefont {Bonetti}}]{127077202}%
	\BibitemOpen
	\bibfield  {author} {\bibinfo {author} {\bibfnamefont {D.}~\bibnamefont
			{Bossini}}, \bibinfo {author} {\bibfnamefont {M.}~\bibnamefont {Pancaldi}},
		\bibinfo {author} {\bibfnamefont {L.}~\bibnamefont {Soumah}}, \bibinfo
		{author} {\bibfnamefont {M.}~\bibnamefont {Basini}}, \bibinfo {author}
		{\bibfnamefont {F.}~\bibnamefont {Mertens}}, \bibinfo {author} {\bibfnamefont
			{M.}~\bibnamefont {Cinchetti}}, \bibinfo {author} {\bibfnamefont
			{T.}~\bibnamefont {Satoh}}, \bibinfo {author} {\bibfnamefont
			{O.}~\bibnamefont {Gomonay}},\ and\ \bibinfo {author} {\bibfnamefont
			{S.}~\bibnamefont {Bonetti}},\ }\bibfield  {title} {\bibinfo {title}
		{Ultrafast amplification and nonlinear magnetoelastic coupling of coherent
			magnon modes in an antiferromagnet},\ }\href
	{https://doi.org/10.1103/PhysRevLett.127.077202} {\bibfield  {journal}
		{\bibinfo  {journal} {Phys. Rev. Lett.}\ }\textbf {\bibinfo {volume} {127}},\
		\bibinfo {pages} {077202} (\bibinfo {year} {2021})}\BibitemShut {NoStop}%
	\bibitem [{\citenamefont {Guo}\ \emph {et~al.}(2025{\natexlab{a}})\citenamefont
		{Guo}, \citenamefont {Wang}, \citenamefont {Wang}, \citenamefont {Zhang},
		\citenamefont {Zhou},\ and\ \citenamefont {Cheng}}]{202505779}%
	\BibitemOpen
	\bibfield  {author} {\bibinfo {author} {\bibfnamefont {Z.}~\bibnamefont
			{Guo}}, \bibinfo {author} {\bibfnamefont {X.}~\bibnamefont {Wang}}, \bibinfo
		{author} {\bibfnamefont {W.}~\bibnamefont {Wang}}, \bibinfo {author}
		{\bibfnamefont {G.}~\bibnamefont {Zhang}}, \bibinfo {author} {\bibfnamefont
			{X.}~\bibnamefont {Zhou}},\ and\ \bibinfo {author} {\bibfnamefont
			{Z.}~\bibnamefont {Cheng}},\ }\bibfield  {title} {\bibinfo {title}
		{Spin-polarized antiferromagnets for spintronics},\ }\href
	{https://doi.org/https://doi.org/10.1002/adma.202505779} {\bibfield
		{journal} {\bibinfo  {journal} {Adv. Mater.}\ }\textbf {\bibinfo {volume}
			{37}},\ \bibinfo {pages} {2505779} (\bibinfo {year}
		{2025}{\natexlab{a}})}\BibitemShut {NoStop}%
	\bibitem [{\citenamefont {Grigorev}\ \emph {et~al.}(2022)\citenamefont
		{Grigorev}, \citenamefont {Filianina}, \citenamefont {Lytvynenko},
		\citenamefont {Sobolev}, \citenamefont {Pokharel}, \citenamefont {Lanz},
		\citenamefont {Sapozhnik}, \citenamefont {Kleibert}, \citenamefont {Bodnar},
		\citenamefont {Grigorev}, \citenamefont {Skourski}, \citenamefont {Kläui},
		\citenamefont {Elmers}, \citenamefont {Jourdan},\ and\ \citenamefont
		{Demsar}}]{Optically}%
	\BibitemOpen
	\bibfield  {author} {\bibinfo {author} {\bibfnamefont {V.}~\bibnamefont
			{Grigorev}}, \bibinfo {author} {\bibfnamefont {M.}~\bibnamefont {Filianina}},
		\bibinfo {author} {\bibfnamefont {Y.}~\bibnamefont {Lytvynenko}}, \bibinfo
		{author} {\bibfnamefont {S.}~\bibnamefont {Sobolev}}, \bibinfo {author}
		{\bibfnamefont {A.~R.}\ \bibnamefont {Pokharel}}, \bibinfo {author}
		{\bibfnamefont {A.~P.}\ \bibnamefont {Lanz}}, \bibinfo {author}
		{\bibfnamefont {A.}~\bibnamefont {Sapozhnik}}, \bibinfo {author}
		{\bibfnamefont {A.}~\bibnamefont {Kleibert}}, \bibinfo {author}
		{\bibfnamefont {S.}~\bibnamefont {Bodnar}}, \bibinfo {author} {\bibfnamefont
			{P.}~\bibnamefont {Grigorev}}, \bibinfo {author} {\bibfnamefont
			{Y.}~\bibnamefont {Skourski}}, \bibinfo {author} {\bibfnamefont
			{M.}~\bibnamefont {Kläui}}, \bibinfo {author} {\bibfnamefont {H.-J.}\
			\bibnamefont {Elmers}}, \bibinfo {author} {\bibfnamefont {M.}~\bibnamefont
			{Jourdan}},\ and\ \bibinfo {author} {\bibfnamefont {J.}~\bibnamefont
			{Demsar}},\ }\bibfield  {title} {\bibinfo {title} {Optically triggered néel
			vector manipulation of a metallic antiferromagnet {M}n$_{2}${A}u under
			strain},\ }\href@noop {} {\bibfield  {journal} {\bibinfo  {journal} {ACS
				Nano}\ }\textbf {\bibinfo {volume} {16}},\ \bibinfo {pages} {20589} (\bibinfo
		{year} {2022})}\BibitemShut {NoStop}%
	\bibitem [{\citenamefont {Jungwirth}\ \emph {et~al.}(2016)\citenamefont
		{Jungwirth}, \citenamefont {Marti}, \citenamefont {Wadley},\ and\
		\citenamefont {Wunderlich}}]{Antif}%
	\BibitemOpen
	\bibfield  {author} {\bibinfo {author} {\bibfnamefont {T.}~\bibnamefont
			{Jungwirth}}, \bibinfo {author} {\bibfnamefont {X.}~\bibnamefont {Marti}},
		\bibinfo {author} {\bibfnamefont {P.}~\bibnamefont {Wadley}},\ and\ \bibinfo
		{author} {\bibfnamefont {J.}~\bibnamefont {Wunderlich}},\ }\bibfield  {title}
	{\bibinfo {title} {Antiferromagnetic spintronics},\ }\href
	{https://doi.org/10.1038/nnano.2016.18} {\bibfield  {journal} {\bibinfo
			{journal} {Nat. Nanotechnol.}\ }\textbf {\bibinfo {volume} {11}},\ \bibinfo
		{pages} {231} (\bibinfo {year} {2016})}\BibitemShut {NoStop}%
	\bibitem [{\citenamefont {Song}\ \emph {et~al.}(2025)\citenamefont {Song},
		\citenamefont {Bai}, \citenamefont {Zhou}, \citenamefont {Han}, \citenamefont
		{Reichlova}, \citenamefont {Dil}, \citenamefont {Liu}, \citenamefont {Chen},\
		and\ \citenamefont {Pan}}]{al1}%
	\BibitemOpen
	\bibfield  {author} {\bibinfo {author} {\bibfnamefont {C.}~\bibnamefont
			{Song}}, \bibinfo {author} {\bibfnamefont {H.}~\bibnamefont {Bai}}, \bibinfo
		{author} {\bibfnamefont {Z.}~\bibnamefont {Zhou}}, \bibinfo {author}
		{\bibfnamefont {L.}~\bibnamefont {Han}}, \bibinfo {author} {\bibfnamefont
			{H.}~\bibnamefont {Reichlova}}, \bibinfo {author} {\bibfnamefont {J.~H.}\
			\bibnamefont {Dil}}, \bibinfo {author} {\bibfnamefont {J.}~\bibnamefont
			{Liu}}, \bibinfo {author} {\bibfnamefont {X.}~\bibnamefont {Chen}},\ and\
		\bibinfo {author} {\bibfnamefont {F.}~\bibnamefont {Pan}},\ }\bibfield
	{title} {\bibinfo {title} {Altermagnets as a new class of functional
			materials},\ }\href {https://doi.org/10.1038/s41578-025-00779-1} {\bibfield
		{journal} {\bibinfo  {journal} {Nat. Rev. Mater.}\ }\textbf {\bibinfo
			{volume} {10}},\ \bibinfo {pages} {473} (\bibinfo {year} {2025})}\BibitemShut
	{NoStop}%
	\bibitem [{\citenamefont {Wang}\ \emph {et~al.}(2025)\citenamefont {Wang},
		\citenamefont {Wang}, \citenamefont {Liu}, \citenamefont {Zhang},\ and\
		\citenamefont {Zhang}}]{al3}%
	\BibitemOpen
	\bibfield  {author} {\bibinfo {author} {\bibfnamefont {D.}~\bibnamefont
			{Wang}}, \bibinfo {author} {\bibfnamefont {H.}~\bibnamefont {Wang}}, \bibinfo
		{author} {\bibfnamefont {L.}~\bibnamefont {Liu}}, \bibinfo {author}
		{\bibfnamefont {J.}~\bibnamefont {Zhang}},\ and\ \bibinfo {author}
		{\bibfnamefont {H.}~\bibnamefont {Zhang}},\ }\bibfield  {title} {\bibinfo
		{title} {Electric-field-induced switchable two-dimensional altermagnets},\
	}\href {https://doi.org/10.1021/acs.nanolett.4c05384} {\bibfield  {journal}
		{\bibinfo  {journal} {Nano Lett.}\ }\textbf {\bibinfo {volume} {25}},\
		\bibinfo {pages} {498} (\bibinfo {year} {2025})}\BibitemShut {NoStop}%
	\bibitem [{\citenamefont {Bai}\ \emph {et~al.}(2024)\citenamefont {Bai},
		\citenamefont {Feng}, \citenamefont {Liu}, \citenamefont {Šmejkal},
		\citenamefont {Mokrousov},\ and\ \citenamefont {Yao}}]{al4}%
	\BibitemOpen
	\bibfield  {author} {\bibinfo {author} {\bibfnamefont {L.}~\bibnamefont
			{Bai}}, \bibinfo {author} {\bibfnamefont {W.}~\bibnamefont {Feng}}, \bibinfo
		{author} {\bibfnamefont {S.}~\bibnamefont {Liu}}, \bibinfo {author}
		{\bibfnamefont {L.}~\bibnamefont {Šmejkal}}, \bibinfo {author}
		{\bibfnamefont {Y.}~\bibnamefont {Mokrousov}},\ and\ \bibinfo {author}
		{\bibfnamefont {Y.}~\bibnamefont {Yao}},\ }\bibfield  {title} {\bibinfo
		{title} {Altermagnetism: Exploring new frontiers in magnetism and
			spintronics},\ }\href
	{https://doi.org/https://doi.org/10.1002/adfm.202409327} {\bibfield
		{journal} {\bibinfo  {journal} {Adv. Funct. Mater.}\ }\textbf {\bibinfo
			{volume} {34}},\ \bibinfo {pages} {2409327} (\bibinfo {year}
		{2024})}\BibitemShut {NoStop}%
	\bibitem [{\citenamefont {Hu}\ \emph {et~al.}(2025)\citenamefont {Hu},
		\citenamefont {Janson}, \citenamefont {Felser}, \citenamefont {McClarty},
		\citenamefont {van~den Brink},\ and\ \citenamefont {G.~Vergniory}}]{al5}%
	\BibitemOpen
	\bibfield  {author} {\bibinfo {author} {\bibfnamefont {M.}~\bibnamefont
			{Hu}}, \bibinfo {author} {\bibfnamefont {O.}~\bibnamefont {Janson}}, \bibinfo
		{author} {\bibfnamefont {C.}~\bibnamefont {Felser}}, \bibinfo {author}
		{\bibfnamefont {P.}~\bibnamefont {McClarty}}, \bibinfo {author}
		{\bibfnamefont {J.}~\bibnamefont {van~den Brink}},\ and\ \bibinfo {author}
		{\bibfnamefont {M.}~\bibnamefont {G.~Vergniory}},\ }\bibfield  {title}
	{\bibinfo {title} {Spin {H}all and {E}delstein effects in chiral
			non-collinear altermagnets},\ }\href
	{https://doi.org/10.1038/s41467-025-64271-8} {\bibfield  {journal} {\bibinfo
			{journal} {Nat. Commun.}\ }\textbf {\bibinfo {volume} {16}},\ \bibinfo
		{pages} {8529} (\bibinfo {year} {2025})}\BibitemShut {NoStop}%
	\bibitem [{\citenamefont {Kawano}\ and\ \citenamefont {Hotta}(2019)}]{thermal}%
	\BibitemOpen
	\bibfield  {author} {\bibinfo {author} {\bibfnamefont {M.}~\bibnamefont
			{Kawano}}\ and\ \bibinfo {author} {\bibfnamefont {C.}~\bibnamefont {Hotta}},\
	}\bibfield  {title} {\bibinfo {title} {Thermal {H}all effect and topological
			edge states in a square-lattice antiferromagnet},\ }\href
	{https://doi.org/10.1103/PhysRevB.99.054422} {\bibfield  {journal} {\bibinfo
			{journal} {Phys. Rev. B}\ }\textbf {\bibinfo {volume} {99}},\ \bibinfo
		{pages} {054422} (\bibinfo {year} {2019})}\BibitemShut {NoStop}%
	\bibitem [{\citenamefont {Guo}\ \emph {et~al.}(2025{\natexlab{b}})\citenamefont
		{Guo}, \citenamefont {Wang}, \citenamefont {Wang}, \citenamefont {Zhang},
		\citenamefont {Zhou},\ and\ \citenamefont {Cheng}}]{spinpolar}%
	\BibitemOpen
	\bibfield  {author} {\bibinfo {author} {\bibfnamefont {Z.}~\bibnamefont
			{Guo}}, \bibinfo {author} {\bibfnamefont {X.}~\bibnamefont {Wang}}, \bibinfo
		{author} {\bibfnamefont {W.}~\bibnamefont {Wang}}, \bibinfo {author}
		{\bibfnamefont {G.}~\bibnamefont {Zhang}}, \bibinfo {author} {\bibfnamefont
			{X.}~\bibnamefont {Zhou}},\ and\ \bibinfo {author} {\bibfnamefont
			{Z.}~\bibnamefont {Cheng}},\ }\bibfield  {title} {\bibinfo {title}
		{Spin-polarized antiferromagnets for spintronics},\ }\href
	{https://doi.org/https://doi.org/10.1002/adma.202505779} {\bibfield
		{journal} {\bibinfo  {journal} {Adv. Mater.}\ }\textbf {\bibinfo {volume}
			{37}},\ \bibinfo {pages} {2505779} (\bibinfo {year}
		{2025}{\natexlab{b}})}\BibitemShut {NoStop}%
	\bibitem [{\citenamefont {Krempaský}\ \emph {et~al.}(2024)\citenamefont
		{Krempaský}, \citenamefont {Šmejkal}, \citenamefont {D’Souza},
		\citenamefont {Hajlaoui}, \citenamefont {Springholz}, \citenamefont
		{Uhlířová}, \citenamefont {Alarab}, \citenamefont {Constantinou},
		\citenamefont {Strocov}, \citenamefont {Usanov}, \citenamefont {Pudelko},
		\citenamefont {González-Hernández}, \citenamefont {Birk~Hellenes},
		\citenamefont {Jansa}, \citenamefont {Reichlová}, \citenamefont {Šobáň},
		\citenamefont {Gonzalez~Betancourt}, \citenamefont {Wadley}, \citenamefont
		{Sinova}, \citenamefont {Kriegner}, \citenamefont {Minár}, \citenamefont
		{Dil},\ and\ \citenamefont {Jungwirth}}]{al2}%
	\BibitemOpen
	\bibfield  {author} {\bibinfo {author} {\bibfnamefont {J.}~\bibnamefont
			{Krempaský}}, \bibinfo {author} {\bibfnamefont {L.}~\bibnamefont
			{Šmejkal}}, \bibinfo {author} {\bibfnamefont {S.~W.}\ \bibnamefont
			{D’Souza}}, \bibinfo {author} {\bibfnamefont {M.}~\bibnamefont {Hajlaoui}},
		\bibinfo {author} {\bibfnamefont {G.}~\bibnamefont {Springholz}}, \bibinfo
		{author} {\bibfnamefont {K.}~\bibnamefont {Uhlířová}}, \bibinfo {author}
		{\bibfnamefont {F.}~\bibnamefont {Alarab}}, \bibinfo {author} {\bibfnamefont
			{P.~C.}\ \bibnamefont {Constantinou}}, \bibinfo {author} {\bibfnamefont
			{V.}~\bibnamefont {Strocov}}, \bibinfo {author} {\bibfnamefont
			{D.}~\bibnamefont {Usanov}}, \bibinfo {author} {\bibfnamefont {W.~R.}\
			\bibnamefont {Pudelko}}, \bibinfo {author} {\bibfnamefont {R.}~\bibnamefont
			{González-Hernández}}, \bibinfo {author} {\bibfnamefont {A.}~\bibnamefont
			{Birk~Hellenes}}, \bibinfo {author} {\bibfnamefont {Z.}~\bibnamefont
			{Jansa}}, \bibinfo {author} {\bibfnamefont {H.}~\bibnamefont {Reichlová}},
		\bibinfo {author} {\bibfnamefont {Z.}~\bibnamefont {Šobáň}}, \bibinfo
		{author} {\bibfnamefont {R.~D.}\ \bibnamefont {Gonzalez~Betancourt}},
		\bibinfo {author} {\bibfnamefont {P.}~\bibnamefont {Wadley}}, \bibinfo
		{author} {\bibfnamefont {J.}~\bibnamefont {Sinova}}, \bibinfo {author}
		{\bibfnamefont {D.}~\bibnamefont {Kriegner}}, \bibinfo {author}
		{\bibfnamefont {J.}~\bibnamefont {Minár}}, \bibinfo {author} {\bibfnamefont
			{J.~H.}\ \bibnamefont {Dil}},\ and\ \bibinfo {author} {\bibfnamefont
			{T.}~\bibnamefont {Jungwirth}},\ }\bibfield  {title} {\bibinfo {title}
		{Altermagnetic lifting of kramers spin degeneracy},\ }\href
	{https://doi.org/10.1038/s41586-023-06907-7} {\bibfield  {journal} {\bibinfo
			{journal} {Nature}\ }\textbf {\bibinfo {volume} {626}},\ \bibinfo {pages}
		{517} (\bibinfo {year} {2024})}\BibitemShut {NoStop}%
	\bibitem [{\citenamefont {\ifmmode~\check{S}\else \v{S}\fi{}mejkal}\ \emph
		{et~al.}(2022{\natexlab{a}})\citenamefont {\ifmmode~\check{S}\else
			\v{S}\fi{}mejkal}, \citenamefont {Sinova},\ and\ \citenamefont
		{Jungwirth}}]{alter3}%
	\BibitemOpen
	\bibfield  {author} {\bibinfo {author} {\bibfnamefont {L.}~\bibnamefont
			{\ifmmode~\check{S}\else \v{S}\fi{}mejkal}}, \bibinfo {author} {\bibfnamefont
			{J.}~\bibnamefont {Sinova}},\ and\ \bibinfo {author} {\bibfnamefont
			{T.}~\bibnamefont {Jungwirth}},\ }\bibfield  {title} {\bibinfo {title}
		{Emerging research landscape of altermagnetism},\ }\href
	{https://doi.org/10.1103/PhysRevX.12.040501} {\bibfield  {journal} {\bibinfo
			{journal} {Phys. Rev. X}\ }\textbf {\bibinfo {volume} {12}},\ \bibinfo
		{pages} {040501} (\bibinfo {year} {2022}{\natexlab{a}})}\BibitemShut
	{NoStop}%
	\bibitem [{\citenamefont {Mazin}(2022)}]{alter4}%
	\BibitemOpen
	\bibfield  {author} {\bibinfo {author} {\bibfnamefont {I.}~\bibnamefont
			{Mazin}} (\bibinfo {collaboration} {The PRX Editors}),\ }\bibfield  {title}
	{\bibinfo {title} {Editorial: Altermagnetism---a new punch line of
			fundamental magnetism},\ }\href {https://doi.org/10.1103/PhysRevX.12.040002}
	{\bibfield  {journal} {\bibinfo  {journal} {Phys. Rev. X}\ }\textbf {\bibinfo
			{volume} {12}},\ \bibinfo {pages} {040002} (\bibinfo {year}
		{2022})}\BibitemShut {NoStop}%
	\bibitem [{\citenamefont {Feng}\ \emph {et~al.}(2022)\citenamefont {Feng},
		\citenamefont {Zhou}, \citenamefont {Šmejkal}, \citenamefont {Wu},
		\citenamefont {Zhu}, \citenamefont {Guo}, \citenamefont
		{González-Hernández}, \citenamefont {Wang}, \citenamefont {Yan},
		\citenamefont {Qin}, \citenamefont {Zhang}, \citenamefont {Wu}, \citenamefont
		{Chen}, \citenamefont {Meng}, \citenamefont {Liu}, \citenamefont {Xia},
		\citenamefont {Sinova}, \citenamefont {Jungwirth},\ and\ \citenamefont
		{Liu}}]{alter5}%
	\BibitemOpen
	\bibfield  {author} {\bibinfo {author} {\bibfnamefont {Z.}~\bibnamefont
			{Feng}}, \bibinfo {author} {\bibfnamefont {X.}~\bibnamefont {Zhou}}, \bibinfo
		{author} {\bibfnamefont {L.}~\bibnamefont {Šmejkal}}, \bibinfo {author}
		{\bibfnamefont {L.}~\bibnamefont {Wu}}, \bibinfo {author} {\bibfnamefont
			{Z.}~\bibnamefont {Zhu}}, \bibinfo {author} {\bibfnamefont {H.}~\bibnamefont
			{Guo}}, \bibinfo {author} {\bibfnamefont {R.}~\bibnamefont
			{González-Hernández}}, \bibinfo {author} {\bibfnamefont {X.}~\bibnamefont
			{Wang}}, \bibinfo {author} {\bibfnamefont {H.}~\bibnamefont {Yan}}, \bibinfo
		{author} {\bibfnamefont {P.}~\bibnamefont {Qin}}, \bibinfo {author}
		{\bibfnamefont {X.}~\bibnamefont {Zhang}}, \bibinfo {author} {\bibfnamefont
			{H.}~\bibnamefont {Wu}}, \bibinfo {author} {\bibfnamefont {H.}~\bibnamefont
			{Chen}}, \bibinfo {author} {\bibfnamefont {Z.}~\bibnamefont {Meng}}, \bibinfo
		{author} {\bibfnamefont {L.}~\bibnamefont {Liu}}, \bibinfo {author}
		{\bibfnamefont {Z.}~\bibnamefont {Xia}}, \bibinfo {author} {\bibfnamefont
			{J.}~\bibnamefont {Sinova}}, \bibinfo {author} {\bibfnamefont
			{T.}~\bibnamefont {Jungwirth}},\ and\ \bibinfo {author} {\bibfnamefont
			{Z.}~\bibnamefont {Liu}},\ }\bibfield  {title} {\bibinfo {title} {An
			anomalous {H}all effect in altermagnetic ruthenium dioxide},\ }\href
	{https://doi.org/10.1038/s41928-022-00866-z} {\bibfield  {journal} {\bibinfo
			{journal} {Nat. Electron.}\ }\textbf {\bibinfo {volume} {5}},\ \bibinfo
		{pages} {735} (\bibinfo {year} {2022})}\BibitemShut {NoStop}%
	\bibitem [{\citenamefont {\ifmmode~\check{S}\else \v{S}\fi{}mejkal}\ \emph
		{et~al.}(2023)\citenamefont {\ifmmode~\check{S}\else \v{S}\fi{}mejkal},
		\citenamefont {Marmodoro}, \citenamefont {Ahn}, \citenamefont
		{Gonz\'alez-Hern\'andez}, \citenamefont {Turek}, \citenamefont {Mankovsky},
		\citenamefont {Ebert}, \citenamefont {D'Souza}, \citenamefont
		{\ifmmode~\check{S}\else \v{S}\fi{}ipr}, \citenamefont {Sinova},\ and\
		\citenamefont {Jungwirth}}]{alter7}%
	\BibitemOpen
	\bibfield  {author} {\bibinfo {author} {\bibfnamefont {L.}~\bibnamefont
			{\ifmmode~\check{S}\else \v{S}\fi{}mejkal}}, \bibinfo {author} {\bibfnamefont
			{A.}~\bibnamefont {Marmodoro}}, \bibinfo {author} {\bibfnamefont {K.-H.}\
			\bibnamefont {Ahn}}, \bibinfo {author} {\bibfnamefont {R.}~\bibnamefont
			{Gonz\'alez-Hern\'andez}}, \bibinfo {author} {\bibfnamefont {I.}~\bibnamefont
			{Turek}}, \bibinfo {author} {\bibfnamefont {S.}~\bibnamefont {Mankovsky}},
		\bibinfo {author} {\bibfnamefont {H.}~\bibnamefont {Ebert}}, \bibinfo
		{author} {\bibfnamefont {S.~W.}\ \bibnamefont {D'Souza}}, \bibinfo {author}
		{\bibfnamefont {O.~c.~v.}\ \bibnamefont {\ifmmode~\check{S}\else
				\v{S}\fi{}ipr}}, \bibinfo {author} {\bibfnamefont {J.}~\bibnamefont
			{Sinova}},\ and\ \bibinfo {author} {\bibfnamefont {T.~c.~v.}\ \bibnamefont
			{Jungwirth}},\ }\bibfield  {title} {\bibinfo {title} {Chiral magnons in
			altermagnetic {R}u{O}$_{2}$},\ }\href
	{https://doi.org/10.1103/PhysRevLett.131.256703} {\bibfield  {journal}
		{\bibinfo  {journal} {Phys. Rev. Lett.}\ }\textbf {\bibinfo {volume} {131}},\
		\bibinfo {pages} {256703} (\bibinfo {year} {2023})}\BibitemShut {NoStop}%
	\bibitem [{\citenamefont {\ifmmode~\check{S}\else \v{S}\fi{}mejkal}\ \emph
		{et~al.}(2022{\natexlab{b}})\citenamefont {\ifmmode~\check{S}\else
			\v{S}\fi{}mejkal}, \citenamefont {Hellenes}, \citenamefont
		{Gonz\'alez-Hern\'andez}, \citenamefont {Sinova},\ and\ \citenamefont
		{Jungwirth}}]{PhysRevX.12.011028}%
	\BibitemOpen
	\bibfield  {author} {\bibinfo {author} {\bibfnamefont {L.}~\bibnamefont
			{\ifmmode~\check{S}\else \v{S}\fi{}mejkal}}, \bibinfo {author} {\bibfnamefont
			{A.~B.}\ \bibnamefont {Hellenes}}, \bibinfo {author} {\bibfnamefont
			{R.}~\bibnamefont {Gonz\'alez-Hern\'andez}}, \bibinfo {author} {\bibfnamefont
			{J.}~\bibnamefont {Sinova}},\ and\ \bibinfo {author} {\bibfnamefont
			{T.}~\bibnamefont {Jungwirth}},\ }\bibfield  {title} {\bibinfo {title} {Giant
			and tunneling magnetoresistance in unconventional collinear antiferromagnets
			with nonrelativistic spin-momentum coupling},\ }\href
	{https://doi.org/10.1103/PhysRevX.12.011028} {\bibfield  {journal} {\bibinfo
			{journal} {Phys. Rev. X}\ }\textbf {\bibinfo {volume} {12}},\ \bibinfo
		{pages} {011028} (\bibinfo {year} {2022}{\natexlab{b}})}\BibitemShut
	{NoStop}%
	\bibitem [{\citenamefont {Zhang}\ \emph
		{et~al.}(2025{\natexlab{a}})\citenamefont {Zhang}, \citenamefont {Cheng},
		\citenamefont {Yin}, \citenamefont {Liu}, \citenamefont {Deng}, \citenamefont
		{Qiao}, \citenamefont {Shi}, \citenamefont {Zhang}, \citenamefont {Lin},
		\citenamefont {Liu}, \citenamefont {Ye}, \citenamefont {Huang}, \citenamefont
		{Meng}, \citenamefont {Zhang}, \citenamefont {Okuda}, \citenamefont
		{Shimada}, \citenamefont {Cui}, \citenamefont {Zhao}, \citenamefont {Cao},
		\citenamefont {Qiao}, \citenamefont {Liu},\ and\ \citenamefont
		{Chen}}]{NP2025cry}%
	\BibitemOpen
	\bibfield  {author} {\bibinfo {author} {\bibfnamefont {F.}~\bibnamefont
			{Zhang}}, \bibinfo {author} {\bibfnamefont {X.}~\bibnamefont {Cheng}},
		\bibinfo {author} {\bibfnamefont {Z.}~\bibnamefont {Yin}}, \bibinfo {author}
		{\bibfnamefont {C.}~\bibnamefont {Liu}}, \bibinfo {author} {\bibfnamefont
			{L.}~\bibnamefont {Deng}}, \bibinfo {author} {\bibfnamefont {Y.}~\bibnamefont
			{Qiao}}, \bibinfo {author} {\bibfnamefont {Z.}~\bibnamefont {Shi}}, \bibinfo
		{author} {\bibfnamefont {S.}~\bibnamefont {Zhang}}, \bibinfo {author}
		{\bibfnamefont {J.}~\bibnamefont {Lin}}, \bibinfo {author} {\bibfnamefont
			{Z.}~\bibnamefont {Liu}}, \bibinfo {author} {\bibfnamefont {M.}~\bibnamefont
			{Ye}}, \bibinfo {author} {\bibfnamefont {Y.}~\bibnamefont {Huang}}, \bibinfo
		{author} {\bibfnamefont {X.}~\bibnamefont {Meng}}, \bibinfo {author}
		{\bibfnamefont {C.}~\bibnamefont {Zhang}}, \bibinfo {author} {\bibfnamefont
			{T.}~\bibnamefont {Okuda}}, \bibinfo {author} {\bibfnamefont
			{K.}~\bibnamefont {Shimada}}, \bibinfo {author} {\bibfnamefont
			{S.}~\bibnamefont {Cui}}, \bibinfo {author} {\bibfnamefont {Y.}~\bibnamefont
			{Zhao}}, \bibinfo {author} {\bibfnamefont {G.-H.}\ \bibnamefont {Cao}},
		\bibinfo {author} {\bibfnamefont {S.}~\bibnamefont {Qiao}}, \bibinfo {author}
		{\bibfnamefont {J.}~\bibnamefont {Liu}},\ and\ \bibinfo {author}
		{\bibfnamefont {C.}~\bibnamefont {Chen}},\ }\bibfield  {title} {\bibinfo
		{title} {Crystal-symmetry-paired spin–valley locking in a layered
			room-temperature metallic altermagnet candidate},\ }\href
	{https://doi.org/10.1038/s41567-025-02864-2} {\bibfield  {journal} {\bibinfo
			{journal} {Nat. Phys.}\ }\textbf {\bibinfo {volume} {21}},\ \bibinfo {pages}
		{760} (\bibinfo {year} {2025}{\natexlab{a}})}\BibitemShut {NoStop}%
	\bibitem [{\citenamefont {Ding}\ \emph {et~al.}(2024)\citenamefont {Ding},
		\citenamefont {Jiang}, \citenamefont {Chen}, \citenamefont {Tao},
		\citenamefont {Liu}, \citenamefont {Li}, \citenamefont {Liu}, \citenamefont
		{Sun}, \citenamefont {Cheng}, \citenamefont {Liu}, \citenamefont {Yang},
		\citenamefont {Zhang}, \citenamefont {Deng}, \citenamefont {Jing},
		\citenamefont {Huang}, \citenamefont {Shi}, \citenamefont {Ye}, \citenamefont
		{Qiao}, \citenamefont {Wang}, \citenamefont {Guo}, \citenamefont {Feng},\
		and\ \citenamefont {Shen}}]{PhysRevLett.133.206401}%
	\BibitemOpen
	\bibfield  {author} {\bibinfo {author} {\bibfnamefont {J.}~\bibnamefont
			{Ding}}, \bibinfo {author} {\bibfnamefont {Z.}~\bibnamefont {Jiang}},
		\bibinfo {author} {\bibfnamefont {X.}~\bibnamefont {Chen}}, \bibinfo {author}
		{\bibfnamefont {Z.}~\bibnamefont {Tao}}, \bibinfo {author} {\bibfnamefont
			{Z.}~\bibnamefont {Liu}}, \bibinfo {author} {\bibfnamefont {T.}~\bibnamefont
			{Li}}, \bibinfo {author} {\bibfnamefont {J.}~\bibnamefont {Liu}}, \bibinfo
		{author} {\bibfnamefont {J.}~\bibnamefont {Sun}}, \bibinfo {author}
		{\bibfnamefont {J.}~\bibnamefont {Cheng}}, \bibinfo {author} {\bibfnamefont
			{J.}~\bibnamefont {Liu}}, \bibinfo {author} {\bibfnamefont {Y.}~\bibnamefont
			{Yang}}, \bibinfo {author} {\bibfnamefont {R.}~\bibnamefont {Zhang}},
		\bibinfo {author} {\bibfnamefont {L.}~\bibnamefont {Deng}}, \bibinfo {author}
		{\bibfnamefont {W.}~\bibnamefont {Jing}}, \bibinfo {author} {\bibfnamefont
			{Y.}~\bibnamefont {Huang}}, \bibinfo {author} {\bibfnamefont
			{Y.}~\bibnamefont {Shi}}, \bibinfo {author} {\bibfnamefont {M.}~\bibnamefont
			{Ye}}, \bibinfo {author} {\bibfnamefont {S.}~\bibnamefont {Qiao}}, \bibinfo
		{author} {\bibfnamefont {Y.}~\bibnamefont {Wang}}, \bibinfo {author}
		{\bibfnamefont {Y.}~\bibnamefont {Guo}}, \bibinfo {author} {\bibfnamefont
			{D.}~\bibnamefont {Feng}},\ and\ \bibinfo {author} {\bibfnamefont
			{D.}~\bibnamefont {Shen}},\ }\bibfield  {title} {\bibinfo {title} {Large band
			splitting in $g$-wave altermagnet {C}r{S}b},\ }\href
	{https://doi.org/10.1103/PhysRevLett.133.206401} {\bibfield  {journal}
		{\bibinfo  {journal} {Phys. Rev. Lett.}\ }\textbf {\bibinfo {volume} {133}},\
		\bibinfo {pages} {206401} (\bibinfo {year} {2024})}\BibitemShut {NoStop}%
	\bibitem [{\citenamefont {Badura}\ \emph {et~al.}(2025)\citenamefont {Badura},
		\citenamefont {Campos}, \citenamefont {Bharadwaj}, \citenamefont {Kounta},
		\citenamefont {Michez}, \citenamefont {Petit}, \citenamefont {Rial},
		\citenamefont {Leiviskä}, \citenamefont {Baltz}, \citenamefont {Krizek},
		\citenamefont {Kriegner}, \citenamefont {Železný}, \citenamefont {Zemen},
		\citenamefont {Telkamp}, \citenamefont {Sailler}, \citenamefont {Lammel},
		\citenamefont {Jaeschke-Ubiergo}, \citenamefont {Hellenes}, \citenamefont
		{González-Hernández}, \citenamefont {Sinova}, \citenamefont {Jungwirth},
		\citenamefont {Goennenwein}, \citenamefont {Šmejkal},\ and\ \citenamefont
		{Reichlova}}]{ex1}%
	\BibitemOpen
	\bibfield  {author} {\bibinfo {author} {\bibfnamefont {A.}~\bibnamefont
			{Badura}}, \bibinfo {author} {\bibfnamefont {W.~H.}\ \bibnamefont {Campos}},
		\bibinfo {author} {\bibfnamefont {V.~K.}\ \bibnamefont {Bharadwaj}}, \bibinfo
		{author} {\bibfnamefont {I.}~\bibnamefont {Kounta}}, \bibinfo {author}
		{\bibfnamefont {L.}~\bibnamefont {Michez}}, \bibinfo {author} {\bibfnamefont
			{M.}~\bibnamefont {Petit}}, \bibinfo {author} {\bibfnamefont
			{J.}~\bibnamefont {Rial}}, \bibinfo {author} {\bibfnamefont {M.}~\bibnamefont
			{Leiviskä}}, \bibinfo {author} {\bibfnamefont {V.}~\bibnamefont {Baltz}},
		\bibinfo {author} {\bibfnamefont {F.}~\bibnamefont {Krizek}}, \bibinfo
		{author} {\bibfnamefont {D.}~\bibnamefont {Kriegner}}, \bibinfo {author}
		{\bibfnamefont {J.}~\bibnamefont {Železný}}, \bibinfo {author}
		{\bibfnamefont {J.}~\bibnamefont {Zemen}}, \bibinfo {author} {\bibfnamefont
			{S.}~\bibnamefont {Telkamp}}, \bibinfo {author} {\bibfnamefont
			{S.}~\bibnamefont {Sailler}}, \bibinfo {author} {\bibfnamefont
			{M.}~\bibnamefont {Lammel}}, \bibinfo {author} {\bibfnamefont
			{R.}~\bibnamefont {Jaeschke-Ubiergo}}, \bibinfo {author} {\bibfnamefont
			{A.~B.}\ \bibnamefont {Hellenes}}, \bibinfo {author} {\bibfnamefont
			{R.}~\bibnamefont {González-Hernández}}, \bibinfo {author} {\bibfnamefont
			{J.}~\bibnamefont {Sinova}}, \bibinfo {author} {\bibfnamefont
			{T.}~\bibnamefont {Jungwirth}}, \bibinfo {author} {\bibfnamefont {S.~T.~B.}\
			\bibnamefont {Goennenwein}}, \bibinfo {author} {\bibfnamefont
			{L.}~\bibnamefont {Šmejkal}},\ and\ \bibinfo {author} {\bibfnamefont
			{H.}~\bibnamefont {Reichlova}},\ }\bibfield  {title} {\bibinfo {title}
		{Observation of the anomalous {N}ernst effect in altermagnetic candidate
			{M}n$_{5}${S}i$_{3}$},\ }\href {https://doi.org/10.1038/s41467-025-62331-7}
	{\bibfield  {journal} {\bibinfo  {journal} {Nat. Commun.}\ }\textbf {\bibinfo
			{volume} {16}},\ \bibinfo {pages} {7111} (\bibinfo {year}
		{2025})}\BibitemShut {NoStop}%
	\bibitem [{\citenamefont {Reimers}\ \emph {et~al.}(2024)\citenamefont
		{Reimers}, \citenamefont {Odenbreit}, \citenamefont {Šmejkal}, \citenamefont
		{Strocov}, \citenamefont {Constantinou}, \citenamefont {Hellenes},
		\citenamefont {Jaeschke~Ubiergo}, \citenamefont {Campos}, \citenamefont
		{Bharadwaj}, \citenamefont {Chakraborty}, \citenamefont {Denneulin},
		\citenamefont {Shi}, \citenamefont {Dunin-Borkowski}, \citenamefont {Das},
		\citenamefont {Kläui}, \citenamefont {Sinova},\ and\ \citenamefont
		{Jourdan}}]{ex2}%
	\BibitemOpen
	\bibfield  {author} {\bibinfo {author} {\bibfnamefont {S.}~\bibnamefont
			{Reimers}}, \bibinfo {author} {\bibfnamefont {L.}~\bibnamefont {Odenbreit}},
		\bibinfo {author} {\bibfnamefont {L.}~\bibnamefont {Šmejkal}}, \bibinfo
		{author} {\bibfnamefont {V.~N.}\ \bibnamefont {Strocov}}, \bibinfo {author}
		{\bibfnamefont {P.}~\bibnamefont {Constantinou}}, \bibinfo {author}
		{\bibfnamefont {A.~B.}\ \bibnamefont {Hellenes}}, \bibinfo {author}
		{\bibfnamefont {R.}~\bibnamefont {Jaeschke~Ubiergo}}, \bibinfo {author}
		{\bibfnamefont {W.~H.}\ \bibnamefont {Campos}}, \bibinfo {author}
		{\bibfnamefont {V.~K.}\ \bibnamefont {Bharadwaj}}, \bibinfo {author}
		{\bibfnamefont {A.}~\bibnamefont {Chakraborty}}, \bibinfo {author}
		{\bibfnamefont {T.}~\bibnamefont {Denneulin}}, \bibinfo {author}
		{\bibfnamefont {W.}~\bibnamefont {Shi}}, \bibinfo {author} {\bibfnamefont
			{R.~E.}\ \bibnamefont {Dunin-Borkowski}}, \bibinfo {author} {\bibfnamefont
			{S.}~\bibnamefont {Das}}, \bibinfo {author} {\bibfnamefont {M.}~\bibnamefont
			{Kläui}}, \bibinfo {author} {\bibfnamefont {J.}~\bibnamefont {Sinova}},\
		and\ \bibinfo {author} {\bibfnamefont {M.}~\bibnamefont {Jourdan}},\
	}\bibfield  {title} {\bibinfo {title} {Direct observation of altermagnetic
			band splitting in {C}r{S}b thin films},\ }\href
	{https://doi.org/10.1038/s41467-024-46476-5} {\bibfield  {journal} {\bibinfo
			{journal} {Nat. Commun.}\ }\textbf {\bibinfo {volume} {15}},\ \bibinfo
		{pages} {2116} (\bibinfo {year} {2024})}\BibitemShut {NoStop}%
	\bibitem [{\citenamefont {He}\ \emph {et~al.}(2024)\citenamefont {He},
		\citenamefont {Li}, \citenamefont {Cui}, \citenamefont {Zhang}, \citenamefont
		{Yu}, \citenamefont {Liu},\ and\ \citenamefont {Zhang}}]{133146602}%
	\BibitemOpen
	\bibfield  {author} {\bibinfo {author} {\bibfnamefont {T.}~\bibnamefont
			{He}}, \bibinfo {author} {\bibfnamefont {L.}~\bibnamefont {Li}}, \bibinfo
		{author} {\bibfnamefont {C.}~\bibnamefont {Cui}}, \bibinfo {author}
		{\bibfnamefont {R.-W.}\ \bibnamefont {Zhang}}, \bibinfo {author}
		{\bibfnamefont {Z.-M.}\ \bibnamefont {Yu}}, \bibinfo {author} {\bibfnamefont
			{G.}~\bibnamefont {Liu}},\ and\ \bibinfo {author} {\bibfnamefont
			{X.}~\bibnamefont {Zhang}},\ }\bibfield  {title} {\bibinfo {title}
		{Quasi-one-dimensional spin transport in altermagnetic ${Z}^{3}$ nodal net
			metals},\ }\href {https://doi.org/10.1103/PhysRevLett.133.146602} {\bibfield
		{journal} {\bibinfo  {journal} {Phys. Rev. Lett.}\ }\textbf {\bibinfo
			{volume} {133}},\ \bibinfo {pages} {146602} (\bibinfo {year}
		{2024})}\BibitemShut {NoStop}%
	\bibitem [{\citenamefont {Gong}\ \emph {et~al.}(2024)\citenamefont {Gong},
		\citenamefont {Wang}, \citenamefont {Han}, \citenamefont {Cheng},
		\citenamefont {Wang}, \citenamefont {Yu},\ and\ \citenamefont
		{Yao}}]{AM36.29}%
	\BibitemOpen
	\bibfield  {author} {\bibinfo {author} {\bibfnamefont {J.}~\bibnamefont
			{Gong}}, \bibinfo {author} {\bibfnamefont {Y.}~\bibnamefont {Wang}}, \bibinfo
		{author} {\bibfnamefont {Y.}~\bibnamefont {Han}}, \bibinfo {author}
		{\bibfnamefont {Z.}~\bibnamefont {Cheng}}, \bibinfo {author} {\bibfnamefont
			{X.}~\bibnamefont {Wang}}, \bibinfo {author} {\bibfnamefont {Z.-M.}\
			\bibnamefont {Yu}},\ and\ \bibinfo {author} {\bibfnamefont {Y.}~\bibnamefont
			{Yao}},\ }\bibfield  {title} {\bibinfo {title} {Hidden real topology and
			unusual magnetoelectric responses in two-dimensional antiferromagnets},\
	}\href {https://doi.org/https://doi.org/10.1002/adma.202402232} {\bibfield
		{journal} {\bibinfo  {journal} {Adv. Mater.}\ }\textbf {\bibinfo {volume}
			{36}},\ \bibinfo {pages} {2402232} (\bibinfo {year} {2024})}\BibitemShut
	{NoStop}%
	\bibitem [{\citenamefont {Gu}\ \emph {et~al.}(2025)\citenamefont {Gu},
		\citenamefont {Liu}, \citenamefont {Zhu}, \citenamefont {Yananose},
		\citenamefont {Chen}, \citenamefont {Hu}, \citenamefont {Stroppa},\ and\
		\citenamefont {Liu}}]{PhysRevLett.134.106802}%
	\BibitemOpen
	\bibfield  {author} {\bibinfo {author} {\bibfnamefont {M.}~\bibnamefont
			{Gu}}, \bibinfo {author} {\bibfnamefont {Y.}~\bibnamefont {Liu}}, \bibinfo
		{author} {\bibfnamefont {H.}~\bibnamefont {Zhu}}, \bibinfo {author}
		{\bibfnamefont {K.}~\bibnamefont {Yananose}}, \bibinfo {author}
		{\bibfnamefont {X.}~\bibnamefont {Chen}}, \bibinfo {author} {\bibfnamefont
			{Y.}~\bibnamefont {Hu}}, \bibinfo {author} {\bibfnamefont {A.}~\bibnamefont
			{Stroppa}},\ and\ \bibinfo {author} {\bibfnamefont {Q.}~\bibnamefont {Liu}},\
	}\bibfield  {title} {\bibinfo {title} {Ferroelectric switchable
			altermagnetism},\ }\href {https://doi.org/10.1103/PhysRevLett.134.106802}
	{\bibfield  {journal} {\bibinfo  {journal} {Phys. Rev. Lett.}\ }\textbf
		{\bibinfo {volume} {134}},\ \bibinfo {pages} {106802} (\bibinfo {year}
		{2025})}\BibitemShut {NoStop}%
	\bibitem [{\citenamefont {Duan}\ \emph {et~al.}(2025)\citenamefont {Duan},
		\citenamefont {Zhang}, \citenamefont {Zhu}, \citenamefont {Liu},
		\citenamefont {Zhang}, \citenamefont {\ifmmode \check{Z}\else
			\v{Z}\fi{}uti\ifmmode~\acute{c}\else \'{c}\fi{}},\ and\ \citenamefont
		{Zhou}}]{PhysRevLett.134.106801}%
	\BibitemOpen
	\bibfield  {author} {\bibinfo {author} {\bibfnamefont {X.}~\bibnamefont
			{Duan}}, \bibinfo {author} {\bibfnamefont {J.}~\bibnamefont {Zhang}},
		\bibinfo {author} {\bibfnamefont {Z.}~\bibnamefont {Zhu}}, \bibinfo {author}
		{\bibfnamefont {Y.}~\bibnamefont {Liu}}, \bibinfo {author} {\bibfnamefont
			{Z.}~\bibnamefont {Zhang}}, \bibinfo {author} {\bibfnamefont
			{I.}~\bibnamefont {\ifmmode \check{Z}\else
				\v{Z}\fi{}uti\ifmmode~\acute{c}\else \'{c}\fi{}}},\ and\ \bibinfo {author}
		{\bibfnamefont {T.}~\bibnamefont {Zhou}},\ }\bibfield  {title} {\bibinfo
		{title} {Antiferroelectric altermagnets: Antiferroelectricity alters
			magnets},\ }\href {https://doi.org/10.1103/PhysRevLett.134.106801} {\bibfield
		{journal} {\bibinfo  {journal} {Phys. Rev. Lett.}\ }\textbf {\bibinfo
			{volume} {134}},\ \bibinfo {pages} {106801} (\bibinfo {year}
		{2025})}\BibitemShut {NoStop}%
	\bibitem [{\citenamefont {Zhang}\ \emph
		{et~al.}(2025{\natexlab{b}})\citenamefont {Zhang}, \citenamefont {Bai},
		\citenamefont {Zou}, \citenamefont {Huang}, \citenamefont {Dai},\ and\
		\citenamefont {Niu}}]{zbbr-426l}%
	\BibitemOpen
	\bibfield  {author} {\bibinfo {author} {\bibfnamefont {Z.}~\bibnamefont
			{Zhang}}, \bibinfo {author} {\bibfnamefont {Y.}~\bibnamefont {Bai}}, \bibinfo
		{author} {\bibfnamefont {X.}~\bibnamefont {Zou}}, \bibinfo {author}
		{\bibfnamefont {B.}~\bibnamefont {Huang}}, \bibinfo {author} {\bibfnamefont
			{Y.}~\bibnamefont {Dai}},\ and\ \bibinfo {author} {\bibfnamefont
			{C.}~\bibnamefont {Niu}},\ }\bibfield  {title} {\bibinfo {title}
		{Altermagnetic quantum spin {H}all effect in a {C}hern homobilayer},\ }\href
	{https://doi.org/10.1103/zbbr-426l} {\bibfield  {journal} {\bibinfo
			{journal} {Phys. Rev. B}\ }\textbf {\bibinfo {volume} {112}},\ \bibinfo
		{pages} {085128} (\bibinfo {year} {2025}{\natexlab{b}})}\BibitemShut
	{NoStop}%
	\bibitem [{\citenamefont {Jin}\ \emph {et~al.}(2025)\citenamefont {Jin},
		\citenamefont {Gong}, \citenamefont {Liu}, \citenamefont {Yang},
		\citenamefont {Zeng}, \citenamefont {Cao},\ and\ \citenamefont
		{Yan}}]{gn6c-1q19}%
	\BibitemOpen
	\bibfield  {author} {\bibinfo {author} {\bibfnamefont {Z.}~\bibnamefont
			{Jin}}, \bibinfo {author} {\bibfnamefont {T.}~\bibnamefont {Gong}}, \bibinfo
		{author} {\bibfnamefont {J.}~\bibnamefont {Liu}}, \bibinfo {author}
		{\bibfnamefont {H.}~\bibnamefont {Yang}}, \bibinfo {author} {\bibfnamefont
			{Z.}~\bibnamefont {Zeng}}, \bibinfo {author} {\bibfnamefont {Y.}~\bibnamefont
			{Cao}},\ and\ \bibinfo {author} {\bibfnamefont {P.}~\bibnamefont {Yan}},\
	}\bibfield  {title} {\bibinfo {title} {Strong coupling of chiral magnons in
			altermagnets},\ }\href {https://doi.org/10.1103/gn6c-1q19} {\bibfield
		{journal} {\bibinfo  {journal} {Phys. Rev. Lett.}\ }\textbf {\bibinfo
			{volume} {135}},\ \bibinfo {pages} {126702} (\bibinfo {year}
		{2025})}\BibitemShut {NoStop}%
	\bibitem [{\citenamefont {Cichutek}\ \emph {et~al.}(2025)\citenamefont
		{Cichutek}, \citenamefont {Kopietz},\ and\ \citenamefont
		{R\"uckriegel}}]{b5vs-ldpm}%
	\BibitemOpen
	\bibfield  {author} {\bibinfo {author} {\bibfnamefont {N.}~\bibnamefont
			{Cichutek}}, \bibinfo {author} {\bibfnamefont {P.}~\bibnamefont {Kopietz}},\
		and\ \bibinfo {author} {\bibfnamefont {A.}~\bibnamefont {R\"uckriegel}},\
	}\bibfield  {title} {\bibinfo {title} {Spontaneous magnon decay in
			two-dimensional altermagnets},\ }\href {https://doi.org/10.1103/b5vs-ldpm}
	{\bibfield  {journal} {\bibinfo  {journal} {Phys. Rev. Res.}\ }\textbf
		{\bibinfo {volume} {7}},\ \bibinfo {pages} {033208} (\bibinfo {year}
		{2025})}\BibitemShut {NoStop}%
	\bibitem [{\citenamefont {Zhang}\ \emph
		{et~al.}(2025{\natexlab{c}})\citenamefont {Zhang}, \citenamefont {Qiu},
		\citenamefont {Chen}, \citenamefont {Wu}, \citenamefont {Wang}, \citenamefont
		{Malik}, \citenamefont {Cai}, \citenamefont {Wu}, \citenamefont {Gao},
		\citenamefont {Hua}, \citenamefont {Yu}, \citenamefont {Xiao}, \citenamefont
		{Jiang}, \citenamefont {Yu}, \citenamefont {Shen},\ and\ \citenamefont
		{Zhang}}]{chiral}%
	\BibitemOpen
	\bibfield  {author} {\bibinfo {author} {\bibfnamefont {Y.}~\bibnamefont
			{Zhang}}, \bibinfo {author} {\bibfnamefont {L.}~\bibnamefont {Qiu}}, \bibinfo
		{author} {\bibfnamefont {J.}~\bibnamefont {Chen}}, \bibinfo {author}
		{\bibfnamefont {S.}~\bibnamefont {Wu}}, \bibinfo {author} {\bibfnamefont
			{H.}~\bibnamefont {Wang}}, \bibinfo {author} {\bibfnamefont {I.~A.}\
			\bibnamefont {Malik}}, \bibinfo {author} {\bibfnamefont {M.}~\bibnamefont
			{Cai}}, \bibinfo {author} {\bibfnamefont {M.}~\bibnamefont {Wu}}, \bibinfo
		{author} {\bibfnamefont {P.}~\bibnamefont {Gao}}, \bibinfo {author}
		{\bibfnamefont {C.}~\bibnamefont {Hua}}, \bibinfo {author} {\bibfnamefont
			{W.}~\bibnamefont {Yu}}, \bibinfo {author} {\bibfnamefont {J.}~\bibnamefont
			{Xiao}}, \bibinfo {author} {\bibfnamefont {Y.}~\bibnamefont {Jiang}},
		\bibinfo {author} {\bibfnamefont {H.}~\bibnamefont {Yu}}, \bibinfo {author}
		{\bibfnamefont {K.}~\bibnamefont {Shen}},\ and\ \bibinfo {author}
		{\bibfnamefont {J.}~\bibnamefont {Zhang}},\ }\bibfield  {title} {\bibinfo
		{title} {Switchable long-distance propagation of chiral magnonic edge
			states},\ }\href {https://doi.org/10.1038/s41563-024-02065-x} {\bibfield
		{journal} {\bibinfo  {journal} {Nat. Mater}\ }\textbf {\bibinfo {volume}
			{24}},\ \bibinfo {pages} {69} (\bibinfo {year}
		{2025}{\natexlab{c}})}\BibitemShut {NoStop}%
	\bibitem [{\citenamefont {Sun}\ \emph {et~al.}(2025)\citenamefont {Sun},
		\citenamefont {Guo}, \citenamefont {Wang}, \citenamefont {Abernathy},
		\citenamefont {Tian},\ and\ \citenamefont {Li}}]{7yhz-jptc}%
	\BibitemOpen
	\bibfield  {author} {\bibinfo {author} {\bibfnamefont {Q.}~\bibnamefont
			{Sun}}, \bibinfo {author} {\bibfnamefont {J.}~\bibnamefont {Guo}}, \bibinfo
		{author} {\bibfnamefont {D.}~\bibnamefont {Wang}}, \bibinfo {author}
		{\bibfnamefont {D.~L.}\ \bibnamefont {Abernathy}}, \bibinfo {author}
		{\bibfnamefont {W.}~\bibnamefont {Tian}},\ and\ \bibinfo {author}
		{\bibfnamefont {C.}~\bibnamefont {Li}},\ }\bibfield  {title} {\bibinfo
		{title} {Observation of chiral magnon band splitting in altermagnetic
			hematite},\ }\href {https://doi.org/10.1103/7yhz-jptc} {\bibfield  {journal}
		{\bibinfo  {journal} {Phys. Rev. Lett.}\ }\textbf {\bibinfo {volume} {135}},\
		\bibinfo {pages} {186703} (\bibinfo {year} {2025})}\BibitemShut {NoStop}%
	\bibitem [{\citenamefont {Haldane}(1988)}]{PhysRevLett.61.2015}%
	\BibitemOpen
	\bibfield  {author} {\bibinfo {author} {\bibfnamefont {F.~D.~M.}\
			\bibnamefont {Haldane}},\ }\bibfield  {title} {\bibinfo {title} {Model for a
			quantum {H}all effect without landau levels: Condensed-matter realization of
			the "parity anomaly"},\ }\href {https://doi.org/10.1103/PhysRevLett.61.2015}
	{\bibfield  {journal} {\bibinfo  {journal} {Phys. Rev. Lett.}\ }\textbf
		{\bibinfo {volume} {61}},\ \bibinfo {pages} {2015} (\bibinfo {year}
		{1988})}\BibitemShut {NoStop}%
	\bibitem [{\citenamefont {\ifmmode~\check{S}\else \v{S}\fi{}abani}\ \emph
		{et~al.}(2020)\citenamefont {\ifmmode~\check{S}\else \v{S}\fi{}abani},
		\citenamefont {Bacaksiz},\ and\ \citenamefont {Milo\ifmmode \check{s}\else
			\v{s}\fi{}evi\ifmmode~\acute{c}\else \'{c}\fi{}}}]{4sm}%
	\BibitemOpen
	\bibfield  {author} {\bibinfo {author} {\bibfnamefont {D.}~\bibnamefont
			{\ifmmode~\check{S}\else \v{S}\fi{}abani}}, \bibinfo {author} {\bibfnamefont
			{C.}~\bibnamefont {Bacaksiz}},\ and\ \bibinfo {author} {\bibfnamefont
			{M.~V.}\ \bibnamefont {Milo\ifmmode \check{s}\else
				\v{s}\fi{}evi\ifmmode~\acute{c}\else \'{c}\fi{}}},\ }\bibfield  {title}
	{\bibinfo {title} {Ab initio methodology for magnetic exchange parameters:
			Generic four-state energy mapping onto a {H}eisenberg spin {H}amiltonian},\
	}\href@noop {} {\bibfield  {journal} {\bibinfo  {journal} {Phys. Rev. B}\
		}\textbf {\bibinfo {volume} {102}},\ \bibinfo {pages} {014457} (\bibinfo
		{year} {2020})}\BibitemShut {NoStop}%
	\bibitem [{\citenamefont {Kresse}\ and\ \citenamefont
		{Furthm\"uller}(1996)}]{vasp}%
	\BibitemOpen
	\bibfield  {author} {\bibinfo {author} {\bibfnamefont {G.}~\bibnamefont
			{Kresse}}\ and\ \bibinfo {author} {\bibfnamefont {J.}~\bibnamefont
			{Furthm\"uller}},\ }\bibfield  {title} {\bibinfo {title} {Efficient iterative
			schemes for ab initio total-energy calculations using a plane-wave basis
			set},\ }\href@noop {} {\bibfield  {journal} {\bibinfo  {journal} {Phys. Rev.
				B}\ }\textbf {\bibinfo {volume} {54}},\ \bibinfo {pages} {11169} (\bibinfo
		{year} {1996})}\BibitemShut {NoStop}%
	\bibitem [{\citenamefont {Perdew}\ \emph {et~al.}(1996)\citenamefont {Perdew},
		\citenamefont {Burke},\ and\ \citenamefont {Ernzerhof}}]{PBE}%
	\BibitemOpen
	\bibfield  {author} {\bibinfo {author} {\bibfnamefont {J.~P.}\ \bibnamefont
			{Perdew}}, \bibinfo {author} {\bibfnamefont {K.}~\bibnamefont {Burke}},\ and\
		\bibinfo {author} {\bibfnamefont {M.}~\bibnamefont {Ernzerhof}},\ }\bibfield
	{title} {\bibinfo {title} {Generalized gradient approximation made simple},\
	}\href@noop {} {\bibfield  {journal} {\bibinfo  {journal} {Phys. Rev. Lett.}\
		}\textbf {\bibinfo {volume} {77}},\ \bibinfo {pages} {3865} (\bibinfo {year}
		{1996})}\BibitemShut {NoStop}%
\end{thebibliography}
\end{document}